%% file: PRE.tex
\documentclass[aps,PRE,superscriptaddress,reprint,longbibliography,floatfix]{revtex4-2}
\input{frontmatter/preamble}

\begin{document}
\title{Probe Dynamics in Active Baths via Trajectory-Response Relations}
\input{frontmatter/authors}

\begin{abstract}
Passive probes in active baths can undergo spontaneous motion driven by nonequilibrium forces that depend on probe velocity and interaction geometry.
Using trajectory-response relations that express the bath response to a probe in terms of free-particle displacement statistics, we investigate the dynamics of freely moving probes weakly coupled to ideal active baths.
The mean force exerted by the bath defines a velocity-space potential whose minima describe the stable states.
By analyzing how the mean force depends on the probe velocity, we identify stable resting and moving states and determine the parameter regime in which they coexist.
For elliptical interactions, the resting state becomes unstable first along the narrow axis, resulting in direction selection already at the linear-response level.
For fourfold interactions, linear friction remains isotropic, while the nonlinear dependence of the mean force on probe velocity selects axial or diagonal motion and produces directional multistability.
We further show that the trajectory-response relations extend to higher orders in the probe strength, where multi-interval free-particle trajectory statistics become relevant.
For a Gaussian probe, these higher-order terms shift the speed of the stable moving state.
\end{abstract}

\maketitle
\input{PRE/01_introduction}
\input{PRE/02_model_probe_dynamics}

\input{PRE/03_speed_selection}

\input{PRE/04_probe_geometry}
\input{PRE/05_higher_order_response}
\input{PRE/06_discussion}
\input{frontmatter/acknowledgments_data}

\appendix
\input{PRE/A_force_integrals}
\input{PRE/B_numerical_methods}
\input{PRE/C_movie_descriptions}
\input{PRE/D_ai_statement}

\bibliography{main}
\end{document}

%% file: frontmatter/preamble.tex
\usepackage{graphicx}
\usepackage{amsmath}
\usepackage{amssymb}
\usepackage{bm}
\usepackage{color}
\usepackage[utf8]{inputenc}
\usepackage[T1]{fontenc}
\usepackage{newunicodechar}
\usepackage{xr-hyper}
\usepackage[colorlinks=true, allcolors=blue]{hyperref}
\usepackage{xcolor}
\usepackage{kotex}
\usepackage{braket}
\usepackage{booktabs}

\newcommand{\dd}{\mathrm d}
\newcommand{\ii}{\mathrm i}
\newcommand{\ee}{\mathrm e}
\newcommand{\avg}[1]{\left\langle #1\right\rangle}
\newcommand{\abs}[1]{\left|#1\right|}
\newcommand{\vq}{\bm q}
\newcommand{\br}{\bm r}
\newcommand{\vV}{\bm V}
\newcommand{\vQ}{\bm Q}

\newcommand{\vu}{\bm u}
\newcommand{\vn}{\bm n}
\newcommand{\vF}{\bm F}
\newcommand{\vR}{\bm R}

\newcommand{\order}{\mathcal O}

%% file: frontmatter/authors.tex
\author{Chul-Ung Woo}
\affiliation{Department of Theoretical Physics and Center for Biophysics, Saarland University, Saarbr\"ucken, Germany}
\author{Jiwon Choi}
\affiliation{Department of Theoretical Physics and Center for Biophysics, Saarland University, Saarbr\"ucken, Germany}
\author{Heiko Rieger}
\affiliation{Department of Theoretical Physics and Center for Biophysics, Saarland University, Saarbr\"ucken, Germany}
\date{\today}

%% file: PRE/01_introduction.tex
\section{Introduction}
\label{sec_I_introduction}

Active matter is driven out of equilibrium by constituents that consume energy to sustain motion or mechanical stresses, including motile microorganisms, motor--filament assemblies, and synthetic self-propelled colloids~\cite{ramaswamy2010mechanics,marchetti2013hydrodynamics,bechinger2016active,bechinger2016active,bowick2022symmetry}.
The interplay of this local driving with interactions gives rise to collective motion~\cite{vicsek1995novel,toner1995long}, phase separation~\cite{tailleur2008statistical,cates2013when,cates2015motility}, and spontaneous flows~\cite{sanchez2012spontaneous}.
Activity also changes how particles accumulate near and exert forces on boundaries and immersed objects, making the mechanical effects of an active medium sensitive to its surroundings~\cite{takatori2014swim,solon2015pressure,nikola2016active,granek2024colloquium}.

Passive probes immersed in active baths provide access to these effects through the forces they experience and the resulting motion~\cite{granek2024colloquium}.
Experiments in microbial suspensions have revealed enhanced diffusion and non-Gaussian displacements~\cite{wu2000particle,leptos2009dynamics,mino2011enhanced,jepson2013enhanced,jeanneret2016entrainment}.
Further studies have demonstrated non-Gaussian transport in motile-cell carpets~\cite{grossmann2024non-gaussian}, transient superdiffusion in bacterial turbulence~\cite{xie2022activity-induced}, and persistent memory in active granular media~\cite{caprini2024emergent}.
Theoretical descriptions connect tracer transport to swimmer--tracer encounters and activity-driven forces~\cite{burkholder2017tracer,dhar2024active}, as well as interaction-induced memory in crowded active suspensions~\cite{reichert2021tracer}.

Active baths can also mediate interactions between passive bodies~\cite{ni2015tunable,baek2018generic,granek2020bodies,ning2023hydrodynamics} and drive directed motion of asymmetric objects~\cite{angelani2009selfstarting,sokolov2009swimming,dileonardo2010bacterial,mallory2014curvature-induced,pietzonka2019autonomous,pellicciotta2025wall}.
To describe driven tracers and probes in these nonequilibrium media, theoretical approaches incorporate friction, noise, and memory through stationary distributions and dynamical correlations conditioned on the probe configuration or motion~\cite{demery2014generalized,maes2015friction,maes2020fluctuating,pei2025induced,pei2026transfer}.
Complementary active-matter theories address effective interactions, mobility, and responses to external perturbations~\cite{steffenoni2016interacting,dalcengio2019linear,burkholder2019fluctuation,solon2022einstein,davis2024active,kafri2026active}.

For externally driven probes, microrheology characterizes how the force exerted by the bath~\cite{reichhardt2015active,burkholder2020nonlinear,peng2022forced,john2025progress}.
A striking response in active baths is negative drag, for which the bath force has a component along the probe velocity and transfers positive mechanical power to the probe~\cite{foffano2012colloids,granek2022anomalous,knippenberg2025negative}.
The velocity dependence of this force can destabilize rest and generate spontaneous motion even for a symmetric passive object in an inversion-symmetric bath~\cite{kim2024symmetry-breaking,pei2026transfer}.
Determining the mean force as a function of velocity is therefore central to identifying resting and moving states and their stability.

We use the trajectory--response relations established in the companion Letter~\cite{companionPRL} to analyze freely translating probe dynamics in active baths.
We first introduce the model and the relations connecting the density perturbation and mean force to the free-particle self-intermediate scattering function (self-ISF).
The resulting mean force describes the quasistatic dynamics of a freely translating probe and defines its velocity-space potential (Sec.~\ref{sec_II_model_probe_dynamics}).
We then use this description to analyze the coexistence of resting and moving states (Sec.~\ref{sec_III_probe_scaling_force}) and the selection of motion directions by interaction geometry (Sec.~\ref{sec_IV_geometry_dependent_motion}).
Extending the response to higher orders in probe strength allows us to examine the correction to the moving speed and connect the higher-order response to correlations between successive displacement intervals (Sec.~\ref{sec_V_higher_order_response}).
We conclude with a discussion in Sec.~\ref{sec_VI_discussion}.

%% file: PRE/02_model_probe_dynamics.tex
\section{Model and probe dynamics}
\label{sec_II_model_probe_dynamics}

\subsection{Stationary density and mean-force response}
\label{sec_II_A_stationary_mean_response}

We summarize the leading trajectory--response relations established in the companion Letter~\cite{companionPRL}.
We consider a dilute ideal gas of noninteracting active particles with far-field number density $\rho_0$.
A passive probe is translated at a constant velocity $\vV$ and interacts with each bath particle through a localized real potential $U(\br)=\epsilon\psi(\br)$, where $\epsilon$ is the probe strength, $\psi$ is a fixed dimensionless profile, and $\br$ denotes the particle position relative to the probe.
The propulsion is represented by a stationary Markov process with internal state $s_t$, active velocity $\vu(s_t)$, stationary distribution $\pi(s)$, and forward generator $\mathcal L_s$.
The stationary distribution satisfies $\mathcal L_s\pi=0$ and $\int\dd s\,\pi(s)=1$.
In the probe frame, the relative particle position obeys
\begin{equation}
\dot{\br}=\vu(s_t)-\vV-\mu\nabla U(\br)
+\sqrt{2D_t}\,\bm\eta(t).
\label{eq_particle_dynamics}
\end{equation}
Here $\mu$ is the particle mobility, $D_t$ is the translational diffusivity, and $\bm\eta$ is a unit Gaussian white noise independent of the propulsion process.
Within this class, the active process need not be Gaussian or isotropic.

Let $\rho(\br;\vV)$ be the stationary particle density in the probe frame, with $\rho(\br;\vV)\to\rho_0$ far from the probe.
Writing $\rho=\rho_0+\delta\rho$ gives
\begin{equation}
\vF_{\rm bath}(\vV)=\int\dd^dr\,\delta\rho(\br;\vV)\nabla U(\br),
\label{eq_force_density_perturbation}
\end{equation}
since $\int\dd^dr\,\nabla U=0$.
We use $f(\br)=\int_{\vq}e^{\ii\vq\cdot\br}f_{\vq}$, with $\int_{\vq}\equiv\int\dd^dq/(2\pi)^d$.
For a free particle initialized with $s_0$ drawn from $\pi$, let $\Delta\br_{0,t}$ be its laboratory-frame displacement.
The self-ISF
$\Phi_0(\vq,t)=\langle e^{-\ii\vq\cdot\Delta\br_{0,t}}\rangle$
is the Fourier transform of its displacement distribution.
Independent translational diffusion gives
$\Phi_0(\vq,t)=e^{-D_tq^2t}\phi_{\rm a}(\vq,t)$, where
$\phi_{\rm a}(\vq,t)=\langle\exp[-\ii\vq\cdot\int_0^t\dd t'\,\vu(s_{t'})]\rangle$.
Define
\begin{equation}
\mathcal G(\vq,\omega)=\int_0^\infty\dd t\,e^{\ii\omega t}\Phi_0(\vq,t).
\label{eq_trajectory_response_transform}
\end{equation}
Uniform translation samples this response at the frequency $\omega_{\vq}=\vq\cdot\vV$.

The retarded density susceptibility and its ideal-bath form are
\begin{align}
\delta\rho_{\vq}(\vV)&=-\chi^R_{\rho U}(\vq,\omega_{\vq})U_{\vq},
\label{eq_density_response_definition}\\
\chi^R_{\rho U}(\vq,\omega)&=\mu\rho_0q^2\mathcal G(\vq,\omega).
\label{eq_density_susceptibility}
\end{align}
Substitution into Eq.~\eqref{eq_force_density_perturbation} gives the leading mean force at order $U^2$:
\begin{equation}
F_{{\rm bath},i}(\vV)
=-\mu\rho_0\int_{\vq}q_iq^2|U_{\vq}|^2
\Im\mathcal G(\vq,\omega_{\vq}).
\label{eq_mean_bath_force}
\end{equation}

For an inversion-symmetric free process, $\Phi_0(\vq,t)$ is real and even in $\vq$.
We use
\begin{equation}
\mathcal K(\vq,\omega)=\frac{\Im\mathcal G(\vq,\omega)}{\omega},
\label{eq_mode_kernel}
\end{equation}
with the zero-frequency value defined by continuity.
Nonnegative $\mathcal K$ at all wave vectors and frequencies excludes positive steady power for every weak probe.
Negative modes can supply positive-power contributions, but their effect on the total force depends on the interaction weight.

\subsection{Free-probe dynamics and stability}
\label{sec_II_B_free_probe_dynamics}

Let $\vR$ be the probe position and $\mathbf x_n$ the position of bath particle $n$.
A freely translating probe of mass $M$ with bare friction $\zeta_0$ obeys
\begin{align}
\dot{\vR}&=\vV,\nonumber\\
M\dot{\vV}&=-\zeta_0\vV+\sum_{n=1}^{N}\nabla U(\mathbf x_n-\vR).
\label{eq_free_probe_microscopic}
\end{align}
Here $V=|\vV|$ is the probe speed.
When the probe velocity varies slowly compared with the relaxation of the bath density profile in the probe frame, we neglect the delayed bath response and approximate the mean probe dynamics using the stationary force in Eq.~\eqref{eq_mean_bath_force}:
\begin{equation}
M\dot{\vV}=\vF_{\rm bath}(\vV)-\zeta_0\vV.
\label{eq_vector_quasistatic_dynamics}
\end{equation}
Uniformly translating states satisfy $\vF_{\rm bath}(\vV_*)-\zeta_0\vV_*=\bm0$.

For an inversion-symmetric bath, the force at small velocity is $F_{{\rm bath},i}=-\gamma_{ij}V_j+O(V^2)$, with
\begin{equation}
\gamma_{ij}=\mu\rho_0\int_{\vq}q_iq_jq^2|U_{\vq}|^2\mathcal K(\vq,0),
\label{eq_linear_friction_tensor}
\end{equation}
where $\mathcal K(\vq,0)=\int_0^\infty\dd t\,t\Phi_0(\vq,t)$.
Resting state is linearly stable when all eigenvalues of $\bm\gamma+\zeta_0\mathsf I$ are positive.
The probe symmetry constrains the force for any imposed velocity.
If an orthogonal transformation $\mathcal R$ leaves the Fourier-space interaction invariant, $\abs{U_{\mathcal R\vq}}^2=\abs{U_{\vq}}^2$, isotropy of the bath implies
\begin{equation}
\vF_{\rm bath}(\mathcal R\vV)=\mathcal R\vF_{\rm bath}(\vV).
\label{eq_force_symmetry}
\end{equation}

For an isotropic probe in an isotropic bath in $d\geq2$, $\vF_{\rm bath}=F_\parallel(V)\widehat{\vV}$, so stable nonzero-speed motion is continuously degenerate in direction.
Perturbations of the speed relax, whereas the direction remains neutral.
In two dimensions, writing the direction of $\vV$ as $\theta$ gives
\begin{equation}
M\dot V=F_\parallel-\zeta_0V,\qquad
MV\dot\theta=F_\perp.
\label{eq_speed_direction_dynamics}
\end{equation}
The longitudinal and transverse forces control speed and direction, respectively.
Anisotropic probe interactions or anisotropic baths can lift this directional degeneracy and select particular directions of motion.

The quasistatic dynamics admits a scalar potential in velocity space,
\begin{equation}
\begin{aligned}
\mathcal W(\vV)={}&\mu\rho_0\int_{\vq}q^2|U_{\vq}|^2
\int_0^{\vq\cdot\vV}\dd\omega\,\Im\mathcal G(\vq,\omega),\\
\mathcal W_{\rm tot}(\vV)={}&\mathcal W(\vV)+\frac{\zeta_0}{2}V^2.
\end{aligned}
\label{eq_velocity_space_potential}
\end{equation}
The right-hand side of Eq.~\eqref{eq_vector_quasistatic_dynamics} is $-\nabla_{\vV}\mathcal W_{\rm tot}$, implying
\begin{equation}
\frac{\dd\mathcal W_{\rm tot}}{\dd t}
=-\frac{1}{M}|\nabla_{\vV}\mathcal W_{\rm tot}|^2\leq0.
\label{eq_potential_decay}
\end{equation}
The linear stability of each fixed point is determined by the Jacobian evaluated at that point.
Stable isolated states correspond to local minima, while saddles separate their basins of attraction.
The potential form of the quasistatic dynamics excludes deterministic limit cycles.
Force fluctuations discussed in the companion Letter~\cite{companionPRL} can broaden the probe velocity distribution around stable states.

\subsection{Active-bath models}
\label{sec_II_C_active_bath_models}

We consider run-and-tumble particles (RTPs) and active Brownian particles (ABPs)~\cite{shaebani2020computational}, parametrized by the two-point velocity correlations
\begin{equation}
\avg{u_i(t)u_j(0)}=\frac{v_0^2}{d}\ee^{-\abs{t}/\tau}\delta_{ij}.
\label{eq_velocity_correlation}
\end{equation}
At fixed dimension $d$, they therefore share the root-mean-square propulsion speed $v_0$, persistence time $\tau$, and long-time active diffusivity $D_a=v_0^2\tau/d$~\cite{cates2013when}.
We set $D_t=0$ for the model-specific results below and use the dimensionless wave number and frequency $z=qv_0\tau$ and $\Omega=\omega\tau$, with $\mathcal K(q,\omega;0)=\tau^2K(z,\Omega)$.

A complete-reset RTP~\cite{tailleur2008statistical,martens2012probability,mori2020universal} moves with $\vu(t)=v_0\vn(t)$, where $\vn$ is a unit vector.
At rate $\alpha=\tau^{-1}$, the propulsion direction is replaced by an independent direction drawn uniformly from $S^{d-1}$.
In one dimension, a complete reset is equivalent to a telegraph process~\cite{angelani2015run-and-tumble,malakar2018steady} with direct sign-flip rate $(2\tau)^{-1}$.

For an ABP~\cite{romanczuk2012active,sevilla2015smoluchowski}, the propulsion velocity is $\vu(t)=v_0\vn(t)$ and the orientation probability density obeys
\begin{equation}
\partial_t p(\vn,t)=D_r\nabla_{S^{d-1}}^2p(\vn,t).
\label{eq_abp_orientation_dynamics}
\end{equation}
The matching in Eq.~\eqref{eq_velocity_correlation} gives $\tau=[(d-1)D_r]^{-1}$.

For these matched models, the companion Letter shows that complete-reset RTPs have nonnegative response for $d\geq2$, whereas one-dimensional RTPs and two-dimensional ABPs possess negative-response modes~\cite{companionPRL}.
Accordingly, the probe-specific calculations below focus on the one-dimensional RTP and the two-dimensional ABP, for which suitable probe weighting can produce negative friction.

For the one-dimensional RTP, the response is 
\begin{equation}
\mathcal G(q,\omega)=\frac{\alpha-\ii\omega}{q^2v_0^2-\omega^2-\ii\alpha\omega}.
\label{eq_rtp1_full_response}
\end{equation}
The corresponding power kernel is
\begin{equation}
K(z,\Omega)=\frac{1+\Omega^2-z^2}{(z^2-\Omega^2)^2+\Omega^2}.
\label{eq_rtp1_exact_kernel}
\end{equation}
The zero-frequency limit is $K(z,0)=(1-z^2)/z^4$.

For the two-dimensional ABP, the angular evolution is represented by a tridiagonal matrix $\mathsf B(z)$ in the normalized cosine basis $\{1,\sqrt{2}\cos\theta,\sqrt{2}\cos2\theta,\ldots\}$, with $l=0$ denoting the constant angular mode. The angular equation is given in Appendix~\ref{app_A_1_abp_response}.

Averaging over the stationary orientation gives
\begin{align}
\phi_{\rm a}(q,t)&=\left[\ee^{-(t/\tau)\mathsf B(z)}\right]_{00},
\label{eq_abp2_isf_matrix}\\
\mathcal G(q,\omega)&=\tau\left[\left(\mathsf B(z)-\ii\Omega\mathsf I\right)^{-1}\right]_{00}.
\label{eq_abp2_response_matrix}
\end{align}
The corresponding continued-fraction evaluation is given in Appendix~\ref{app_B_1_numerical_theory}.

%% file: PRE/03_speed_selection.tex
\section{Speed selection and coexistence}
\label{sec_III_probe_scaling_force}

\subsection{Gaussian probes and velocity-dependent mean force}
\label{sec_III_A_finite_speed_response}

The interaction spectrum $\abs{U_{\vq}}^2$ sets the wave-vector weighting of the bath response characterized in Sec.~\ref{sec_II_C_active_bath_models}.
We first consider a radial potential with a single length scale, $U(\br)=\epsilon\,\psi(\br/a)$, where $\epsilon$ sets the potential amplitude, $a$ is its characteristic width, and $\psi$ is a fixed dimensionless profile.
For an isotropic bath and radial probe, the bath force is parallel to $\vV$.
We write $\vF_{\rm bath}=F_\parallel(V)\widehat{\vV}$ and $\gamma(V)=-F_\parallel(V)/V$ for $V=|\vV|>0$, with $\gamma(0)$ defined by continuity.
Choosing the polar axis along $\vV$, let $\varphi=\angle(\vq,\vV)$.
Equation~\eqref{eq_mean_bath_force} gives
\begin{equation}
F_\parallel(V)=-\mu\rho_0V\int_{\vq}q^4\abs{U_{\vq}}^2\cos^2\varphi\,\mathcal K(q,qV\cos\varphi;D_t).
\label{eq_longitudinal_force}
\end{equation}
At fixed $\psi$, changing $a$ changes the wave-vector weighting, while $\epsilon^2$ multiplies the entire leading force. In particular, the force-balance speed at $\zeta_0=0$ is independent of $\epsilon$ at this order.

We specialize to the isotropic repulsive Gaussian potential $U_g(r)=\epsilon\exp[-r^2/(2a^2)]$, with $\epsilon>0$.
Set $k=aq$, $\Gamma=v_0\tau/a$, and $\nu=V/v_0$, and denote the area of $S^{d-1}$ by $\mathcal A_{d-1}$.
Using $U_{g,\vq}=(2\pi)^{d/2}\epsilon a^d\ee^{-a^2q^2/2}$, the linear friction at $D_t=0$ is
\begin{equation}
\frac{\gamma(0)}{\rho_0}=\frac{\mathcal A_{d-1}}{d}\mu\epsilon^2\tau^2a^{d-4}\int_0^\infty\dd k\,k^{d+3}\ee^{-k^2}K(k\Gamma,0).
\label{eq_gaussian_linear_friction}
\end{equation}
For the one-dimensional RTP with the Gaussian potential,
\begin{equation}
\frac{\gamma(0)}{\rho_0}=\frac{\sqrt\pi\mu\epsilon^2}{av_0^2}\left(\Gamma^{-2}-\frac12\right).
\label{eq_rtp_linear_friction}
\end{equation}
The friction changes sign at $\Gamma_c=\sqrt{2}$.
For the two-dimensional ABP, numerical evaluation of Eq.~\eqref{eq_gaussian_linear_friction} gives a single zero at $\Gamma_c\simeq1.78$, with positive friction below and negative friction above. 
Representative linear-friction and density-field comparisons with particle simulations are given in the companion Letter~\cite{companionPRL}.

For the one-dimensional RTP, the leading finite-speed force can be evaluated from Eqs.~\eqref{eq_mean_bath_force} and \eqref{eq_rtp1_exact_kernel}.
For a Gaussian potential, the result is
\begin{equation}
\begin{aligned}
\frac{F_\parallel(V)}{\rho_0}=-\frac{2\mu\epsilon^2}{av_0}\,\nu & \int_0^\infty \dd k \,k^2\ee^{-k^2}\\
&\times\frac{1-\Gamma^2(1-\nu^2)k^2}{\nu^2+\Gamma^2(1-\nu^2)^2k^2}.
\end{aligned}
\label{eq_rtp_gaussian_force_integral}
\end{equation}
The force is odd in the signed velocity $V$, and the integral remains finite at $\abs{V}=v_0$.

\begin{figure}[tbp]
\centering
\includegraphics[width=\columnwidth]{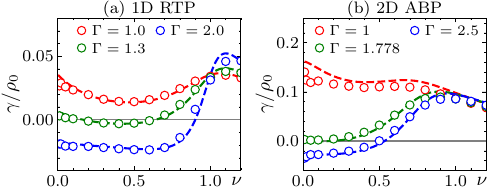}
\caption{\textbf{Velocity-dependent friction of Gaussian probes.}
Effective friction divided by the bath density as a function of the reduced imposed speed $\nu=V/v_0$ for (a) a one-dimensional RTP bath and (b) a two-dimensional ABP bath.
Dashed curves show the weak-potential theory and open symbols show fixed-velocity simulations.
The representative values of $\Gamma$ span regimes of positive and negative linear friction; the horizontal line marks $\gamma=0$.}
\label{fig_1_gaussian_friction}
\end{figure}

For a two-dimensional ABP, the finite-speed response is obtained by evaluating the angular hierarchy at the frequencies $\Omega=\nu z\cos\varphi$ within the Gaussian-probe integral.
Figure~\ref{fig_1_gaussian_friction} shows the resulting effective friction for the one-dimensional RTP and two-dimensional ABP together with the fixed-velocity simulations used in the companion Letter~\cite{companionPRL}.

\begin{figure*}[t]
\centering
\includegraphics[width=\linewidth]{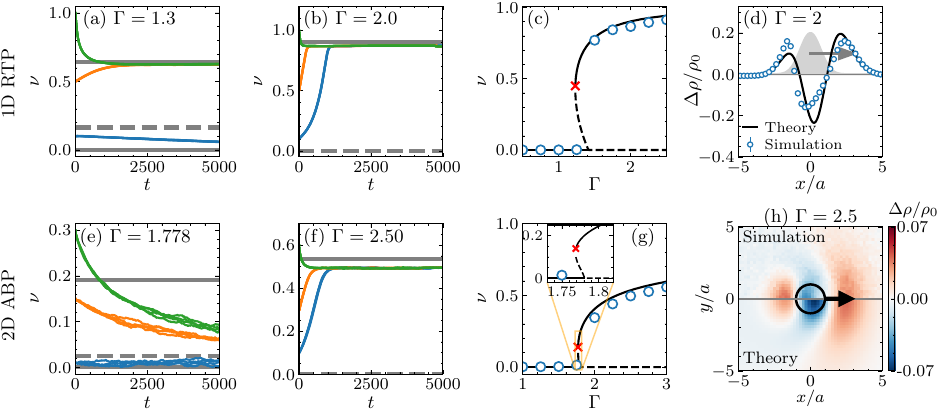}
\caption{\textbf{Dynamics and force-balance states of freely moving probes.}
The upper row (a)-(d) shows the one-dimensional RTP, and the lower row (e)-(h) the two-dimensional ABP.
(a,b,e,f) Time evolution of the normalized probe speed $\nu$ from several initial conditions, with (a,e) in the coexistence regime $\Gamma_{\rm sn}<\Gamma<\Gamma_c$ and (b,f) above $\Gamma_c$.
(c,g) Positive force-balance roots versus $\Gamma$; open circles show long-time speeds from freely translating probe simulations.
Throughout, solid and dashed lines denote states or branches stable and unstable with respect to speed perturbations, respectively.
Red crosses mark saddle-node points where the stable and unstable moving branches meet; the inset in (g) enlarges the ABP bifurcation region.
(d,h) Normalized density difference $\Delta\rho(\mathbf r, \nu^\ast)/\rho_0$ around freely moving probe with steady-state speed $\nu^\ast$. Shaded region in (d) shows a schematic Gaussian potential and black circle shows the width $a=1$ of the probe. 
Arrows indicate the direction of steady probe motion.
See also Supplemental Movie 1~\cite{sm}.}
\label{fig_2_free_probe_dynamics}
\end{figure*}

\subsection{Low- and high-speed limits}
\label{sec_III_B_speed_limits}

At small speed, consider an isotropic free-particle process with finite long-time diffusivity $D$.
At long wavelengths on diffusive time scales,
\begin{equation}
\mathcal G(\vq,\omega)\simeq\frac{1}{Dq^2-\ii\omega}.
\label{eq_diffusive_response}
\end{equation}
For a localized real probe with $U_0\equiv U_{\vq=0}\neq0$, the long-wavelength contribution gives the following force corrections. Their derivation is given in Appendix~\ref{app_A_3_speed_limits}.

In one dimension, with $V$ interpreted as the signed velocity, the force is
\begin{equation}
\frac{F_\parallel(V)}{\rho_0}=-\frac{\gamma(0)}{\rho_0}V+\frac{\mu|U_0|^2}{2D^3}V|V|+O(V^3).
\label{eq_diffusive_1d_force}
\end{equation}

In two dimensions, the force is

\begin{equation}
\begin{aligned}
\frac{F_\parallel(V)}{\rho_0}={}&-\frac{\gamma(0)}{\rho_0}V+\frac{3\mu|U_0|^2}{16\pi D^4}V^3\ln\!\left(\frac{V_0}{|V|}\right)+O(V^3).
\end{aligned}
\label{eq_diffusive_2d_force}
\end{equation}
Here $V_0$ includes the finite terms from wave numbers beyond the diffusive regime.

For $d>2$, the coefficient of the leading nonlinear correction from the diffusive long-wavelength mode is infrared convergent and the corresponding force correction is analytic.

For the one-dimensional RTP with the Gaussian probe, the force has the nonanalytic small-speed expansion
\begin{equation}
\begin{aligned}
\frac{F_\parallel(V)}{\rho_0}=\frac{\mu\epsilon^2}{av_0}\biggl[&-\sqrt{\pi}\left(\Gamma^{-2}-\frac{1}{2}\right)\nu\\
&+\frac{\pi}{\Gamma^3}\nu\abs{\nu}+\order(\nu^3)\biggr].
\end{aligned}
\label{eq_rtp_small_speed_force}
\end{equation}
The linear term agrees with Eq.~\eqref{eq_rtp_linear_friction}, and the $\nu|\nu|$ term agrees with Eq.~\eqref{eq_diffusive_1d_force} for $D=v_0^2\tau$.
When negative-response modes dominate at finite speed, a propulsive interval can develop while rest remains stable. Its boundaries follow from the full force.

At sufficiently high speed, the bath force becomes resistive and provides saturation of the motion.
For a sufficiently regular localized potential, the large-speed force is
\begin{equation}
\frac{F_\parallel(V)}{\rho_0}=-\frac{\mu}{V}\int\dd^dr\,|\nabla U(\br)|^2+o(V^{-1}).
\label{eq_large_speed_force}
\end{equation}
The high-speed result follows from the short-time property $\Phi_0(\vq,0)=1$.
The probe crosses the interaction region on a time scale proportional to $a/V$, so the leading force samples the short-time displacement statistics. Appendix~\ref{app_A_3_speed_limits} gives the derivation.

\subsection{Resting and moving states}
\label{sec_III_C_free_probe_dynamics}

The speed follows the radial equation in Eq.~\eqref{eq_speed_direction_dynamics}.
The resting state is linearly stable when $\gamma(0)>0$ and unstable when $\gamma(0)<0$ for $\zeta_0=0$.
For $\gamma(0)<0$, the propulsive small-speed force and the resistive high-speed limit in Eq.~\eqref{eq_large_speed_force} guarantee at least one nonzero solution of $F_\parallel(V_*)=0$, provided the force is continuous.
The radial dynamics gives a stable simple root when $F_\parallel'(V_*)<0$.
At a nonzero root, this condition is equivalent to $\gamma'(V_*)>0$.
Figure~\ref{fig_1_gaussian_friction}(a,b) illustrates how the friction changes with persistence in the RTP and ABP models.
At $\Gamma=1$, the friction remains positive in both models.
At intermediate persistence ($\Gamma=1.3$ for RTPs and $\Gamma=1.778$ for ABP), the curves develop a negative-friction interval despite $\gamma(0)>0$.
At larger persistence ($\Gamma=2$ for RTPs and $\Gamma=2.5$ for ABPs), the friction is negative at small speeds but becomes positive at higher speeds, yielding stable moving roots.

The freely translating trajectories illustrate the distinction between the coexistence regime and the regime above $\Gamma_c$.
In the one-dimensional RTP coexistence regime, trajectories initialized on opposite sides of the unstable root relax to rest or to the stable moving branch [Fig.~\ref{fig_2_free_probe_dynamics}(a)].
For the two-dimensional ABP, this separation is less clear near coexistence~[Fig.~\ref{fig_2_free_probe_dynamics}(e)].

In both the RTP and ABP, stable and unstable moving branches appear at a model-dependent saddle-node $\Gamma_{\rm sn}<\Gamma_c$, and the stable branch continues through $\Gamma_c$~[Fig.~\ref{fig_2_free_probe_dynamics}(c) and (g)].
For $\Gamma_{\rm sn}<\Gamma<\Gamma_c$, a linearly stable resting state coexists with a stable moving state, separated by an unstable moving root.
In this regime, small velocity perturbations relaxes to rest, whereas a sufficiently large initial speed leads to the stable moving state.
In the velocity-space potential, the coexisting stable states correspond to distinct local minima, whose values need not be equal.
Within the quasistatic radial dynamics, this branch structure can also permit hysteresis under parameter variation.

Figures~\ref{fig_2_free_probe_dynamics}(d,h) show the density differences $\Delta\rho_{\nu^\ast}/\rho_0$ relative to the resting-probe profiles at the selected moving speeds.
Although the profiles are not strictly symmetric, their contributions to the bath force from opposite sides of the probe cancel at the selected speed, consistent with $F_\parallel(V_*)=0$.
The theoretical profiles quantitatively reproduce the spatial features observed in simulations.

Above $\Gamma_c$, the RTP speed approaches the force-balance root closely, whereas the mean ABP speed remains slightly below the constant-velocity value [Figs.~\ref{fig_2_free_probe_dynamics}(c) and (g)].
We return to the lower ABP speed in Sec.~\ref{sec_V_A_finite_amplitude_probe}, where the corrections in probe strength shift the moving root downward.

%% file: PRE/04_probe_geometry.tex
\begin{figure}[t]
\centering
\includegraphics[width=\columnwidth]{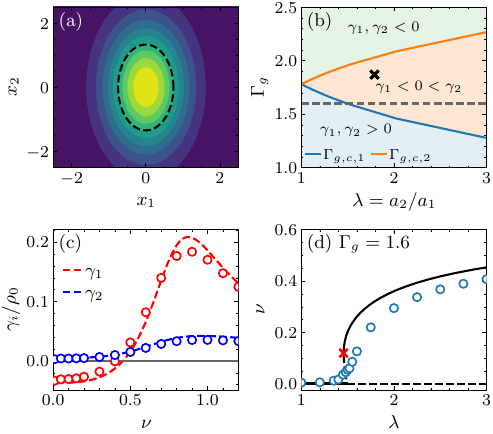}
\caption{
\textbf{Directional selection by an elliptical Gaussian probe.}
(a) Probe geometry with $a_1<a_2$ at fixed $a_g=\sqrt{a_1a_2}=1$. The black dashed contour indicates the ellipse corresponding to the aspect ratio $\lambda=a_2/a_1$ marked by the $\times$ in panel (b). 
(b) Linear-response state diagram in the $(\lambda,\Gamma_g)$ plane; the critical curves mark $\gamma_1=0$ and $\gamma_2=0$, and the $\times$ denotes the parameters used in panel (c).
The gray dashed line corresponds to the parameters in (d).
(c) Effective friction $\gamma_i(V)=-F_i(V)/V$, divided by the bath density, along the two principal axes.
The dashed lines denote the theoretical prediction of each axis and open circles show simulation
(d) Positive force-balance roots along the narrow axis as functions of $\lambda$ at fixed $\Gamma_g=1.6$.
The solid and dashed nonzero branches are stable and unstable against speed perturbations.
Open circles show the simulated free-probe speeds, and the red cross marks the saddle-node.
}
\label{fig_3_elliptical_probe_selection}
\end{figure}

\begin{figure*}[t]
\centering
\includegraphics[width=0.99\textwidth]{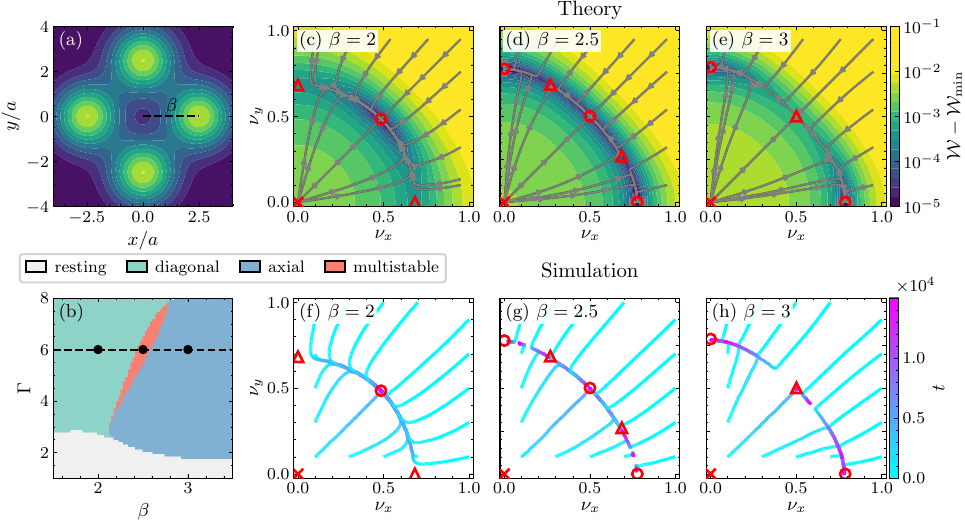}
\caption{
\textbf{Nonlinear directional selection and multistability of a fourfold probe.}
(a) Fourfold Gaussian potential with dimensionless offset $\beta=b/a$.
(b) Nonlinear state diagram in the $(\beta,\Gamma)$ plane.
Colors denote resting, diagonal, axial, and multistable regimes, and the horizontal dashed line marks the cut $\Gamma=6$ used in panels (c)--(h).
The probe orientation is fixed.
(c)--(e) Theoretical velocity-space flows at $\Gamma=6$ for $\beta=2$, $2.5$, and $3$, respectively.
Open circles denote stable moving fixed points, triangles denote saddle points, and crosses denote the unstable fixed points.
Background colormaps represent the scalar potential of velocity $\mathcal W-\mathcal W_{\min}$ for a fourfold probe. 
(f)--(h) Corresponding velocity-space trajectories from simulations of freely moving probes.
We use the reduced velocity vector $\bm\nu=\vV/v_0$.
Colors in panels (f)--(h) indicate the simulation time $t$.
}
\label{fig_4_fourfold_probe_multistability}
\end{figure*}

\section{Geometry-dependent direction selection}
\label{sec_IV_geometry_dependent_motion}

Anisotropic interactions can select preferred directions of probe motion.
We keep the same isotropic two-dimensional ABP bath and vary only $U$, changing the wave-vector weighting without changing $\mathcal K(q,\omega)$.
Anisotropic passive tracers have previously been used to probe shape-dependent transport in active suspensions~\cite{peng2016diffusion,yang2016dynamics,nordanger2022anisotropic}.
Here we consider two inversion-symmetric probe geometries: an elliptical Gaussian probe and a probe with fourfold symmetry. 
Because of the symmetries of both probes, each moving state has a counterpart with the opposite velocity, rather than a preferred direction of rectified motion~\cite{nikola2016active,pietzonka2019autonomous,lee2021geometry,olsonreichhardt2017ratchet}.
Throughout this section, the probe orientation is fixed in the laboratory frame.
We consider both the stationary force at imposed velocity and the dynamics of a freely translating probe.

\subsection{Elliptical probe: linear direction selection}
\label{sec_IV_A_elliptical_probe}

As a minimal example of geometry-dependent response, we consider an elliptical Gaussian probe with fixed body axes $\widehat{\bm e}_1$ and $\widehat{\bm e}_2$,
\begin{equation}
U_{\rm ell}(x_1,x_2) 
= \epsilon \exp\left( -\frac{x_1^2}{2a_1^2} -\frac{x_2^2}{2a_2^2} \right),
\label{eq_elliptical_probe}
\end{equation}
where $x_i=\br\cdot\widehat{\bm e}_i$ and $a_1<a_2$.
We characterize the probe by the geometric-mean width
$a_g=\sqrt{a_1a_2}$, the aspect ratio $\lambda=a_2/a_1$, and the persistence ratio $\Gamma_g=v_0\tau/a_g$.

At fixed $a_g$, the integrated interaction strength is $\int\dd^2r\,U_{\rm ell}=2\pi\epsilon a_g^2$. Changing $\lambda$ redistributes the wave-vector weight at fixed integrated strength.
The narrower real-space direction has a broader Fourier weight, so the two principal-axis frictions sample different wave-vector contributions in Eq.~\eqref{eq_linear_friction_tensor}.

For motion along a body axis, define $\gamma_i(V)=-F_i(V)/V$.
Appendix~\ref{app_A_force_integrals} gives the evaluation integral.
Reflection symmetry about the two body axes makes them principal directions of the linear friction, $\bm\gamma=\operatorname{diag}(\gamma_1,\gamma_2)$ in the body frame.
The anisotropic interaction splits the circular-probe threshold into the critical curves $\Gamma_{g,c,1}(\lambda)$ and $\Gamma_{g,c,2}(\lambda)$, defined by $\gamma_i(0)=0$.
For $\Gamma_{g,c,1}(\lambda)<\Gamma_g<\Gamma_{g,c,2}(\lambda)$, the linear friction satisfies $\gamma_1<0<\gamma_2$.
Thus the resting state is unstable only along the narrow axis, and this one-axis instability region broadens with increasing aspect ratio [Fig.~\ref{fig_3_elliptical_probe_selection}(b)].

Reflection symmetry ensures that the transverse bath force vanishes when the probe moves along either principal axis.
The zeros of the narrow-axis force give symmetry-related nonzero force-balance states $\vV=\pm V\widehat{\bm e}_1$, with stable and unstable speed branches.
When $\gamma_1<0$, the change from a propulsive force at small speed to a resistive force at high speed guarantees a nonzero narrow-axis root.
The critical curves locate the loss of stability of rest; stability of a finite-speed state also requires a restoring transverse force.
Fixed-velocity simulations reproduce the calculated friction along both axes [Fig.~\ref{fig_3_elliptical_probe_selection}(c)], while freely translating probes follow the stable narrow-axis branch [Fig.~\ref{fig_3_elliptical_probe_selection}(d) and Supplemental Movie 2~\cite{sm}].

\subsection{Fourfold probe: nonlinear direction selection and multistability}
\label{sec_IV_B_fourfold_probe}

\begin{figure}
\centering
\includegraphics{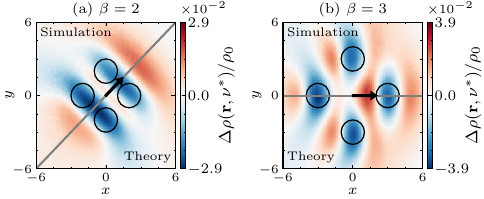}
\caption{
\textbf{Density field around moving fourfold Gaussian probe.}
Normalized density difference $\Delta\rho(\mathbf r, {\bm \nu}^\ast)/\rho_0$ around a fourfold Gaussian probe moving with steady-state velocity $\bm\nu^\ast$.
Panels (a) and (b) show $\beta=2$ and $\beta=3$, corresponding to diagonal and axial motion, respectively.
Black circles represent four Gaussian probe components and the black arrow indicates the direction of probe motion.
Both panels use $\Gamma=6$, with all other parameters as in Fig.~\ref{fig_4_fourfold_probe_multistability}.
}
\label{fig_clover_density_field}
\end{figure}

The fourfold probe is constructed by superposing four copies of the Gaussian potential defined in Sec.~\ref{sec_III_probe_scaling_force}, centered at displacements $\pm b\widehat{\bm x}$ and $\pm b\widehat{\bm y}$:
\begin{equation}
U_4(\br)=\frac{1}{4}\sum_{\sigma=\pm1}
\left[U_g(\br-\sigma b\widehat{\bm x})+U_g(\br-\sigma b\widehat{\bm y})\right].
\label{eq_fourfold_probe}
\end{equation}
The normalization recovers the single Gaussian probe at $b=0$, and we use $\beta=b/a$.
In Fourier space, $U_{4,\vq}=U_{g,\vq}[\cos(q_xb)+\cos(q_yb)]/2$; the corresponding force quadrature is given in Appendix~\ref{app_A_force_integrals}.

For this interaction, Eq.~\eqref{eq_force_symmetry} with $\mathcal R=\mathcal R_{\pi/2}$ implies fourfold rotational symmetry of the force.
Together with the symmetry of the linear friction tensor, this makes the linear friction isotropic, so directional selection must arise from the nonlinear velocity dependence of the force.
As $\beta$ increases, the selected motion changes from diagonal to axial through a regime of directional multistability [Fig.~\ref{fig_4_fourfold_probe_multistability}(b)].

We determine the stability of each moving state from the velocity-space potential defined in Eq.~\eqref{eq_velocity_space_potential}.
Since $\vF_{\rm bath}=-\nabla_{\vV}\mathcal W$, we characterize this shape using the dimensionless force Jacobian
\begin{equation}
J_{ij}=\frac{av_0}{\mu\rho_0\epsilon^2\tau}
\left.\frac{\partial F_{{\rm bath},i}}{\partial V_j}\right|_{\vV=\vV_*}.
\label{eq_fourfold_jacobian}
\end{equation}
At each axial or diagonal root, the Jacobian has eigenvectors parallel and perpendicular to $\vV_*$, with eigenvalues $j_\parallel$ and $j_\perp$, respectively.
When both eigenvalues are negative, the potential rises under small changes in either speed or direction, so the root is a stable local minimum.
When $j_\parallel j_\perp<0$, the potential rises in one direction and falls in the other, identifying a saddle.

Along the representative cut $\Gamma=6$, the potential has diagonal minima at $\beta=2$, both diagonal and axial minima at $\beta=2.5$, and axial minima at $\beta=3$ [Fig.~\ref{fig_4_fourfold_probe_multistability}(c)-(e)].
The superposed force streamlines show relaxation toward these minima.
In the multistable regime, four diagonal and four axial minima coexist with eight oblique saddles whose stable manifolds separate the basins of axial and diagonal motion.

Freely translating probes reproduce the diagonal, multistable, and axial regimes predicted by the stationary force [Fig.~\ref{fig_4_fourfold_probe_multistability}(f--h) and Supplemental Movies 3--5~\cite{sm}].
At $\beta=2.5$, probes initialized near diagonal and axial attractors remain in distinct long-lived moving states, providing a direct dynamical signature of directional multistability.
Figure~\ref{fig_clover_density_field} shows the steady-state density differences around freely translating fourfold probes at $\Gamma=6$.
The profiles accompany diagonal motion at $\beta=2$ and axial motion at $\beta=3$, providing a real-space view of the directions selected by the free-probe dynamics.
In both cases, the probe orientation remains fixed while the surrounding bath density adjusts to its motion.

%% file: PRE/05_higher_order_response.tex
\section{Finite-strength corrections and multi-interval trajectory statistics}
\label{sec_V_higher_order_response}

Beyond the quadratic force, increasing probe strength changes both the selected speed and the trajectory information entering the response: successive perturbations couple displacement intervals along the same free trajectory.
The lower ABP speed in Fig.~\ref{fig_2_free_probe_dynamics}(f,g) provides a motivation for evaluating these corrections.
We first compute their effect on the Gaussian-probe speed and then express the higher-order response through multi-interval trajectory statistics.
Let $P(\br,s,t)$ denote the number density resolved with respect to the propulsion state.
Its evolution under Eq.~\eqref{eq_particle_dynamics} is
\begin{equation}
\begin{aligned}
\partial_tP={}&-[\vu(s)-\vV]\cdot\nabla P+D_t\nabla^2P+\mathcal L_sP\\
&+\mu\nabla\cdot[P\nabla U].
\end{aligned}
\label{eq_joint_evolution}
\end{equation}
The free generator is
\begin{equation}
\mathcal L_0=-[\vu(s)-\vV]\cdot\nabla+D_t\nabla^2+\mathcal L_s.
\label{eq_free_evolution_operator}
\end{equation}
We expand the joint density in the probe potential as
\begin{equation}
P(\br,s,t)=\rho_0\pi(s)+\sum_{n=1}^{\infty}\delta P_n(\br,s,t).
\label{eq_density_expansion}
\end{equation}
Here $\delta P_n=\order(U^n)$, and we define $\delta P_0=\rho_0\pi(s)$.
For $n\geq1$, collecting equal powers of $U$ in Eq.~\eqref{eq_joint_evolution} gives
\begin{equation}
\partial_t\delta P_n=\mathcal L_0\delta P_n
+\mu\nabla\cdot\left[\delta P_{n-1}\nabla U\right].
\label{eq_density_recursion}
\end{equation}

For the stationary response, we set $\partial_t\delta P_n=0$ in Eq.~\eqref{eq_density_recursion} and solve successively in $n$.
With $U=\epsilon\psi$, density corrections through $n=p-1$ give the force through order $\epsilon^p$,
\begin{equation}
\vF_{\rm bath}^{[p]}(\vV,\epsilon)
=\sum_{n=1}^{p-1}\int\dd^dr\,\dd s\,
\delta P_n(\br,s;\vV)\nabla U(\br).
\label{eq_truncated_bath_force}
\end{equation}
\subsection{Finite-strength Gaussian probe}
\label{sec_V_A_finite_amplitude_probe}

For the circular Gaussian probe of Sec.~\ref{sec_III_probe_scaling_force}, we evaluate finite-strength corrections in a two-dimensional ideal ABP bath at $\Gamma=2.5$ and $D_t=0$.
The stationary recursion in Eq.~\eqref{eq_density_recursion} is solved in a spatial Fourier and angular basis, as detailed in Appendix~\ref{app_B_numerical_methods}.
For translation along $x$, we denote the force by $F_x^{[p]}(\nu,\epsilon)$ and the corresponding effective friction by $\gamma^{[p]}(V,\epsilon)=-F_x^{[p]}(V/v_0,\epsilon)/V$.
The positive-speed force-balance root satisfies
\begin{equation}
F_x^{[p]}(\nu_*^{[p]},\epsilon)=0.
\label{eq_finite_amplitude_root}
\end{equation}

\begin{figure}[t]
\centering
\includegraphics[width=\columnwidth]{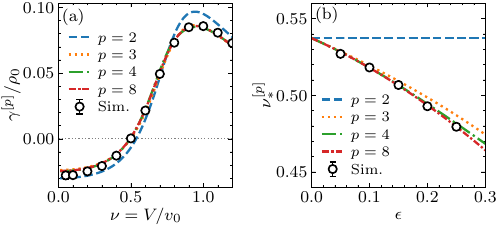}
\caption{\textbf{Higher-order corrections for a Gaussian probe in a two-dimensional ABP bath.}
(a) Effective friction $\gamma^{[p]}/\rho_0$ as a function of $\nu=V/v_0$ at $\epsilon=0.2$.
(b) Positive-speed force-balance roots $\nu_*^{[p]}$ as the probe strength is varied.
Curves retain force terms through order $\epsilon^p$ for $p=2,3,4,8$; open circles show particle simulations.
Both panels use $\Gamma=2.5$ and $D_t=0$.
The $p=2$ reference in (b) uses the same periodic box as the higher-order curves.}
\label{fig_5_finite_amplitude_response}
\end{figure}

Figure~\ref{fig_5_finite_amplitude_response}(a) shows that higher-order corrections bring the effective friction closer to simulations while preserving the crossover from propulsive to resistive forcing.
At the leading-order moving root, these corrections make the friction positive, corresponding to a negative correction to the mean force; since the friction has a positive slope at its stable zero, the moving root shifts downward.
In panel (b), the higher-order roots decrease with probe amplitude and reproduce the probe speeds observed in simulations, improving the leading-order prediction used in Fig.~\ref{fig_2_free_probe_dynamics}(f).

This shift can be quantified by expanding about a simple force zero.
For motion along a fixed symmetry axis, write $F_x/(\rho_0\epsilon^2)=c_2(\nu)+\epsilon c_3(\nu)+O(\epsilon^2)$, with the bath and probe profile fixed.
If $c_2(\nu_0)=0$ and $c_2'(\nu_0)\neq0$, the moving root is
\begin{equation}
\nu_*(\epsilon)=\nu_0-\epsilon\frac{c_3(\nu_0)}{c_2'(\nu_0)}+O(\epsilon^2).
\label{eq_moving_root_correction}
\end{equation}
At a stable speed root, $c_2'(\nu_0)<0$.
The expansion of a simple root does not apply at a saddle-node, where this derivative vanishes.

\subsection{Multi-interval trajectory statistics}
\label{sec_V_B_multi_interval_statistics}

Successive perturbations by the probe act along the trajectory of a single bath particle. Its propulsion state evolves continuously between them, so the intervening displacement intervals are generally correlated. Their joint statistics enter the higher-order response even when different bath particles are independent.

For an initially unperturbed bath, $\delta P_n(0)=0$ for $n\geq1$, and the formal solution of Eq.~\eqref{eq_density_recursion} is
\begin{equation}
\delta P_n(t)=\mu\int_0^t\dd t'\,
\ee^{(t-t')\mathcal L_0}
\nabla\cdot\left[\delta P_{n-1}(t')\nabla U\right].
\label{eq_higher_order_formal_solution}
\end{equation}
To iterate this relation in Fourier space, each probe interaction contributes the convolution
\begin{equation}
[\nabla\cdot(f\nabla U)]_{\vq}
=-\int_{\vq'}[\vq\cdot(\vq-\vq')]U_{\vq-\vq'}f_{\vq'}.
\label{eq_higher_order_fourier_convolution}
\end{equation}
Between successive probe interactions, a mode with wave vector $\vq$ propagates under
\begin{equation}
\mathcal L_0(\vq)
=\mathcal L_s-D_tq^2-\ii\vq\cdot[\vu(s)-\vV].
\label{eq_higher_order_fourier_generator}
\end{equation}

Repeated application of Eq.~\eqref{eq_density_recursion} introduces consecutive free-propagation intervals of durations $t_1,\ldots,t_n$.
In Fourier space, let $\vq_j$ be the wave vector supplied by the $j$th probe interaction and define $\vQ_j=\sum_{a=1}^{j}\vq_a$.
The $j$th free-propagation interval therefore carries the cumulative wave vector $\vQ_j$, whose mode is sampled by uniform probe motion at $\omega_{\vQ_j}$.

In the stationary limit, the second-order response is

\begin{widetext}
\begin{equation}
\begin{aligned}
\delta P_{2,\vq}
=\mu^2\rho_0\int_{\vq_1}\int_{\vq_2}
(2\pi)^d\delta^{(d)}(\vq-\vQ_2)\,
q_1^2U_{\vq_1}(\vQ_2\cdot\vq_2)U_{\vq_2}
\int_0^\infty\dd t_2\int_0^\infty\dd t_1\,
\ee^{t_2\mathcal L_0(\vQ_2)}
\ee^{t_1\mathcal L_0(\vQ_1)}\pi .
\end{aligned}
\label{eq_higher_order_second_order}
\end{equation}

Let $T_j=\sum_{\ell=1}^{j}t_\ell$, with $T_0=0$.
The multi-interval displacement statistics are
\begin{equation}
\Phi_0^{(n)}(\vQ_1,\ldots,\vQ_n;t_1,\ldots,t_n)
=\left\langle
\exp\left[-\ii\sum_{j=1}^{n}\vQ_j\cdot
\left(\Delta\br_{0,T_j}-\Delta\br_{0,T_{j-1}}\right)\right]
\right\rangle .
\label{eq_multi_interval_characteristic_function}
\end{equation}
The average is over a single free trajectory initialized from $\pi(s)$.

Integrating the ordered propagators over the final propulsion state gives
\begin{equation}
\begin{aligned}
\int\dd s\,
\ee^{t_n\mathcal L_0(\vQ_n)}\cdots
\ee^{t_2\mathcal L_0(\vQ_2)}
\ee^{t_1\mathcal L_0(\vQ_1)}\pi(s)
=\exp\left[\ii\sum_{j=1}^{n}\omega_{\vQ_j}t_j\right]
\Phi_0^{(n)}(\vQ_1,\ldots,\vQ_n;t_1,\ldots,t_n).
\end{aligned}
\label{eq_higher_order_ordered_propagators}
\end{equation}

The corresponding frequency transform is
\begin{equation}
\begin{aligned}
\mathcal G^{(n)}(\vQ_1,\ldots,\vQ_n;\omega_1,\ldots,\omega_n)
=\int_0^\infty\dd t_1\cdots\int_0^\infty\dd t_n\,
\ee^{\ii\sum_{j=1}^{n}\omega_jt_j}
\Phi_0^{(n)}(\vQ_1,\ldots,\vQ_n;t_1,\ldots,t_n).
\end{aligned}
\label{eq_multi_interval_response_transform}
\end{equation}

Writing $\delta\rho_n=\int\dd s\,\delta P_n$, iteration gives the stationary density response at order $U^n$,
\begin{equation}
\begin{aligned}
\delta\rho_{n,\vq}
=(-\mu)^n\rho_0
\int_{\vq_1}\cdots\int_{\vq_n}
(2\pi)^d\delta^{(d)}(\vq-\vQ_n)\,
q_1^2U_{\vq_1}
\prod_{j=2}^{n}
\left[(\vQ_j\cdot\vq_j)U_{\vq_j}\right]
\mathcal G^{(n)}
(\vQ_1,\ldots,\vQ_n;
\omega_{\vQ_1},\ldots,\omega_{\vQ_n}).
\end{aligned}
\label{eq_higher_density_general}
\end{equation}
\end{widetext}

The force in Eq.~\eqref{eq_truncated_bath_force} can then be evaluated as
\begin{equation}
F_{{\rm bath},i}^{[p]}(\vV,\epsilon)
=-\ii\sum_{n=1}^{p-1}\int_{\vq}q_iU_{-\vq}\,\delta\rho_{n,\vq}.
\label{eq_higher_force_general}
\end{equation}
For $p=2$, $\mathcal G^{(1)}$ reduces to $\mathcal G$, and Eqs.~\eqref{eq_higher_density_general} and~\eqref{eq_higher_force_general} recover the leading density and force responses of Sec.~\ref{sec_II_A_stationary_mean_response}.
For $n=2$, related multi-time extensions of the self-ISF have been used to characterize correlations between successive particle displacements in supercooled liquids~\cite{kim2010multitime}.
Higher-order trajectory statistics have also been calculated for active Brownian trajectories~\cite{squarcini2022spectral}.

The weight $|U_{\vq}|^2$ used in Secs.~\ref{sec_III_probe_scaling_force} and~\ref{sec_IV_geometry_dependent_motion} leaves the leading force insensitive to the Fourier phase and to polarity reversal, $U(\br)\to U(-\br)$.
The cubic correction contains products of the form $U_{\vq}U_{\vq'}U_{-\vq-\vq'}$.
These products retain Fourier-phase information that can distinguish $U(\br)$ from $U(-\br)$.
The inversion-symmetric geometries in Sec.~\ref{sec_IV_geometry_dependent_motion} preserve opposite motion directions at every order. The phase dependence above permits a polarity dependence for asymmetric interactions.

%% file: PRE/06_discussion.tex
\section{Conclusion and discussion}
\label{sec_VI_discussion}

We have used trajectory--response relations to connect free-particle displacement statistics to the velocity-dependent mean force and the resulting dynamics of passive probes in ideal active baths.
For Gaussian probes in one-dimensional RTP and two-dimensional ABP baths, the quasistatic analysis identifies a regime where stable resting and moving states coexist, bounded by a saddle-node and the loss of linear stability of rest.
Probe geometry determines how motion directions are selected: elliptical interactions produce anisotropic linear friction, whereas fourfold symmetry leaves linear friction isotropic and requires nonlinear velocity dependence to select axial or diagonal motion and produce directional multistability.
The velocity-space potential organizes these states as minima and saddles.
Comparisons with freely translating probe simulations support the predicted moving branches and geometry-dependent direction selection.

For a Gaussian probe in a two-dimensional ABP bath, higher-order corrections shift the selected moving speed downward as the probe strength increases over the range studied, improving agreement with simulations.
Higher orders in probe strength involve correlations between successive free-particle displacement intervals and retain Fourier-phase information that is absent from the leading force.

The quasistatic approximation does not account for the delayed adjustment of the bath to a changing probe velocity.
For probe motion varying on the bath-relaxation timescale, memory effects may become important~\cite{knezevic2021oscillatory}.
The present theory also does not address interacting baths, where collective dynamics and nonlinear density structure require a different treatment~\cite{granek2020bodies,granek2024colloquium}.
For probes with multiple stable directions of motion, how fluctuations induce transitions between these states and affect long-time transport remains an open question~\cite{kim2024symmetry-breaking,pei2026transfer}.

%% file: frontmatter/acknowledgments_data.tex
\section*{acknowledgments}
We thank Jae Dong Noh for helpful discussions.  


%% file: PRE/A_force_integrals.tex
\section{Bath response and probe-force integrals}
\label{app_A_force_integrals}

\subsection{Two-dimensional ABP response}
\label{app_A_1_abp_response}

For the two-dimensional ABP, let $\theta$ denote the angle between $\vn$ and $\vq$ and write $\bar t=t/\tau$. 
We define the characteristic function of the active displacement conditioned on the initial orientation by
\begin{equation}
f(\theta,\bar t)=\left\langle\exp\left[-\ii\vq\cdot\int_0^t\dd t'\,\vu(t')\right]\middle|\theta(0)=\theta\right\rangle .    
\label{eq_abp2_conditional_characteristic}
\end{equation}
It satisfies
\begin{equation}
\begin{aligned}
\partial_{\bar t}f(\theta,\bar t)&=\partial_\theta^2f(\theta,\bar t)-\ii z\cos\theta\,f(\theta,\bar t)\\
&\equiv-\mathsf B(z)f(\theta,\bar t),
\end{aligned}
\label{eq_abp2_angular_evolution}
\end{equation}
with $f(\theta,0)=1$. 
Averaging the angular propagator gives Eqs.~\eqref{eq_abp2_isf_matrix} and~\eqref{eq_abp2_response_matrix}. The continued fraction used to evaluate this matrix element is given in Appendix~\ref{app_B_1_numerical_theory}.

\subsection{Low- and high-speed limits}
\label{app_A_3_speed_limits}

The diffusive approximation in Eq.~\eqref{eq_diffusive_response} gives the nonanalytic terms retained in Sec.~\ref{sec_III_B_speed_limits}.
The corresponding power kernel is
\begin{equation}
\mathcal K(\vq,\omega)\simeq\frac{1}{D^2q^4+\omega^2}.
\label{eq_diffusive_power_kernel}
\end{equation}
For a localized real probe with $U_0\equiv U_{\vq=0}\neq0$, subtracting the zero-speed friction in Eq.~\eqref{eq_mean_bath_force} gives the small-$q$ contribution
\begin{equation}
\begin{aligned}
\frac{\gamma(V)-\gamma(0)}{\rho_0}\simeq{}&-\frac{\mu\abs{U_0}^2V^2}{D^2}\int_{\abs{\vq}<q_{\rm uv}}\frac{\dd^dq}{(2\pi)^d}\\
&\times\frac{(\widehat{\vq}\cdot\widehat{\vV})^4}{D^2q^2+V^2(\widehat{\vq}\cdot\widehat{\vV})^2},
\end{aligned}
\label{eq_diffusive_smallq_general}
\end{equation}
where $q_{\rm uv}$ denotes the upper wave-number scale over which the diffusive approximation applies.

In one dimension, taking $V$ as the signed velocity along the chosen axis gives
\begin{equation}
\begin{aligned}
\int_{-q_{\rm uv}}^{q_{\rm uv}}\frac{\dd q}{2\pi}\frac{1}{D^2q^2+V^2}&=\frac{1}{\pi D\abs V}\tan^{-1}\left(\frac{Dq_{\rm uv}}{\abs V}\right)\\
&=\frac{1}{2D\abs V}+\order(1).
\end{aligned}
\label{eq_diffusive_1d_integral}
\end{equation}
This yields Eq.~\eqref{eq_diffusive_1d_force}.

In two dimensions, Eq.~\eqref{eq_diffusive_smallq_general} contains the radial integral
\begin{equation}
\int_0^{q_{\rm uv}}\frac{\dd q\,q}{D^2q^2+V^2\cos^2\varphi}=\frac{1}{2D^2}\ln\left[\frac{D^2q_{\rm uv}^2+V^2\cos^2\varphi}{V^2\cos^2\varphi}\right].
\label{eq_diffusive_2d_integral}
\end{equation}
For $\abs V\ll Dq_{\rm uv}$, the term proportional to $\ln\abs V$ is obtained using $\int_0^{2\pi}\dd\varphi\,\cos^4\varphi=3\pi/4$.
The remaining finite terms can be absorbed into a velocity scale $V_0$, giving Eq.~\eqref{eq_diffusive_2d_force}.

To derive the high-speed asymptotic force, choose the $x$ axis along the imposed velocity, $\vV=V\widehat{\bm x}$.
Define
\begin{equation}
\mathcal B(\br)=\int_{\vq}q^2\abs{U_{\vq}}^2\ee^{\ii\vq\cdot\br}.
\label{eq_large_speed_probe_kernel}
\end{equation}
Using Eq.~\eqref{eq_mean_bath_force}, the longitudinal force can then be written as 
\begin{equation}
\frac{F_\parallel(V)}{\rho_0}=\frac{\mu}{V}\int_0^\infty\dd y\,\left\langle\partial_x\mathcal B \left(y\,\widehat{\bm x}-\Delta\br_{0,y/V}\right)\right\rangle,
\label{eq_large_speed_rescaled_integral}
\end{equation}
with $y=Vt$.
For fixed $y$, the free-particle displacement satisfies $\Delta\br_{0,y/V}\to\bm0$ as $V\to\infty$.
For a sufficiently regular localized potential, the short-time limit can therefore be taken inside the integral, giving
\begin{equation}
\frac{F_\parallel(V)}{\rho_0}=\frac{\mu}{V}\int_0^\infty\dd y\,\partial_x\mathcal B(y\widehat{\bm x})+o(V^{-1}).
\label{eq_large_speed_asymptotic_integral}
\end{equation}
Since $\mathcal B(\br)\to0$ as $\abs{\br}\to\infty$, this gives Eq.~\eqref{eq_large_speed_force}, with $\mathcal B(\bm0)=\int_{\vq}q^2|U_{\vq}|^2=\int\dd^dr\,|\nabla U|^2$ by Parseval's identity.

\subsection{Anisotropic probes}

For the elliptical Gaussian probe in Eq.~\eqref{eq_elliptical_probe}, write $\widehat{\vq}=\cos\vartheta\,\widehat{\bm e}_1+\sin\vartheta\,\widehat{\bm e}_2$.
With $z=qv_0\tau$ and $\nu=V/v_0$, the friction $\gamma_i(V)=-F_i(V)/V$ for motion along $\widehat{\bm e}_i$ is
\begin{equation}
\begin{aligned}
\frac{\gamma_i(V)}{\rho_0}
={}&\frac{\mu\epsilon^2\tau^2}{a_g^2\Gamma_g^6}
\int_0^{2\pi}\dd\vartheta\,(\widehat{\vq}\cdot\widehat{\bm e}_i)^2
\int_0^\infty\dd z\,z^5 K \left(z,\Omega\right)\\
&\times\exp \left[-\frac{z^2}{\Gamma_g^2}
\left(\lambda^{-1}\cos^2\vartheta+\lambda\sin^2\vartheta\right)\right].
\end{aligned}
\label{eq_elliptical_friction}
\end{equation}

For the fourfold probe in Eq.~\eqref{eq_fourfold_probe}, using $k=aq$, $\widehat{\vq}=(\cos\vartheta,\sin\vartheta)$, and $\bm\nu=\vV/v_0$, the dimensionless force becomes
\begin{equation}
\begin{aligned}
\frac{a\vF_{\rm bath}(v_0\bm\nu)}{\mu\rho_0\epsilon^2\tau}&
=-\frac{\Gamma}{4}\int_0^\infty\dd k\,k^5\ee^{-k^2} \\
&\times\int_0^{2\pi}\dd\vartheta\,\widehat{\vq}(\widehat{\vq}\cdot\bm\nu) K \left(z,\Omega\right)\\
&\times\left[\cos(k\beta\cos\vartheta)+\cos(k\beta\sin\vartheta)\right]^2 .
\end{aligned}
\label{eq_fourfold_dimensionless_force}
\end{equation}
Differentiating the dimensionless force with respect to $\nu_j$ gives the Jacobian in Eq.~\eqref{eq_fourfold_jacobian}.

%% file: PRE/B_numerical_methods.tex
\section{Numerical methods and simulations}
\label{app_B_numerical_methods}

Throughout the numerical calculations and simulations, we set $D_t=0$ and use dimensionless units with $a=v_0=\mu=1$.
Finite-box calculations use $L=32$ for the one-dimensional RTP and $L=16$ for the two-dimensional ABP.

\subsection{Numerical evaluation of the response theory}
\label{app_B_1_numerical_theory}

For the leading-order ABP response, we evaluate the matrix element in Eq.~\eqref{eq_abp2_response_matrix} using a backward continued fraction.
In the normalized cosine basis, the diagonal elements of $\mathsf B$ are $l^2$, and its off-diagonal elements are $\ii zc_l$, with $c_0=1/\sqrt{2}$ and $c_l=1/2$ for $l\geq1$.
Truncating at $l_{\max}$, we initialize $R_{l_{\max}}=l_{\max}^2+\delta z^2-\ii\Omega$ and recurse downward using
\begin{equation}
R_l=l^2+\delta z^2-\ii\Omega+\frac{z^2c_l^2}{R_{l+1}},
\label{eq_abp_continued_fraction}
\end{equation}
where $\delta=D_t/(v_0^2\tau)$.
The response is $\mathcal G(q,\omega)=\tau/R_0$, and the force kernel follows from Eq.~\eqref{eq_mode_kernel}, with its zero-frequency value defined by continuity.

Probe forces are evaluated using Eq.~\eqref{eq_mean_bath_force}, with the Gaussian scaling integral in Sec.~\ref{sec_III_probe_scaling_force} and the anisotropic quadratures in Appendix~\ref{app_A_force_integrals}.
Elliptical force-balance branches are obtained from the force along the principal axes, while fourfold moving states are obtained from the full two-dimensional force and classified using both eigenvalues of its Jacobian.
We use $l_{\max}=50$ and a relative convergence target of $10^{-6}$ for nonzero observables under refinement of the truncation order, quadrature resolution, and integration cutoffs.

For the circular Gaussian potential $U=\epsilon\psi$, with $\psi(\br)=\exp(-r^2/2)$, the stationary density is expanded in powers of $\epsilon$ and angular Fourier modes as
\begin{equation}
P(\br,\theta)=\frac{\rho_0}{2\pi}\left[1+\sum_{n\geq1}\epsilon^n\sum_m f_{n,m}(\br)e^{\ii m\theta}\right].
\label{eq_amplitude_density_modes}
\end{equation}
For $\vV=\nu\widehat{\bm x}$, the stationary recursion in Eq.~\eqref{eq_density_recursion} gives, in spatial Fourier space,
\begin{equation}
\begin{aligned}
&\left(\frac{m^2}{\Gamma}-\ii q_x\nu\right)f_{n,m}(\vq)
+\frac{\ii}{2}(q_x-\ii q_y)f_{n,m-1}(\vq)\\
&\quad+\frac{\ii}{2}(q_x+\ii q_y)f_{n,m+1}(\vq)
=\left[\nabla\cdot(f_{n-1,m}\nabla\psi)\right]_{\vq},
\end{aligned}
\end{equation}
where the right-hand side denotes a spatial Fourier transform and $f_{0,m}=\delta_{m0}$.
The first-order source is $-q^2\psi_{\vq}$ for $m=0$ and zero otherwise.
At each subsequent order, the source is constructed from the preceding density correction, evaluating spatial derivatives spectrally and products in real space.
For each $\vq$, the resulting tridiagonal linear system is solved over the angular modes $-16\leq m\leq16$ to obtain $f_{n,m}(\vq)$.

On the $64\times64$ spatial grid with spacing $\Delta x=L/64$, the force coefficients are evaluated as
\begin{align}
\frac{F_x^{[p]}(\nu,\epsilon)}{\rho_0\epsilon^2}
&=\sum_{j=2}^{p}\epsilon^{j-2}c_j(\nu),\\
c_j(\nu)
&=\Delta x^2\sum_{x,y} f_{j-1,0}(x,y)\,\partial_x\psi(x,y).
\end{align}

\subsection{Particle simulations}
\label{app_B_2_particle_simulations}

The simulations evolve noninteracting bath particles in periodic boxes according to Eq.~\eqref{eq_particle_dynamics}, with $N=10^7$ particles in both models and $N=\rho_0L^d$.
Particle positions are advanced with an Euler--Maruyama time step $\Delta t=0.01$.
For the ABP, the orientation angle undergoes rotational diffusion over the same time step, while the RTP propulsion sign is updated with probability $p_{\rm flip}=[1-\exp(-\Delta t/\tau)]/2$.
The probe strength is $\epsilon=0.2$, except for the amplitude scan in Fig.~\ref{fig_5_finite_amplitude_response}(b).

For imposed-velocity simulations, the probe position is prescribed, and the bath force is obtained by summing the microscopic probe forces over all particles and averaging over the stationary measurement interval.
For freely translating probes, Eq.~\eqref{eq_particle_dynamics} and the probe dynamics in Eq.~\eqref{eq_free_probe_microscopic} are integrated simultaneously, with the imposed velocity replaced by the instantaneous probe velocity.
Unless stated otherwise, we set $M=\rho_0\tilde M$ with $\tilde M=10$ and $\zeta_0=0$, and include no independent noise or propulsion of the probe.
For anisotropic probes, the probe orientation is held fixed.

%% file: PRE/C_movie_descriptions.tex
\section{Description of supplemental movies}

All movies show passive probes in two-dimensional ABP baths and use $L=16$, $v_0=1$, $\mu=1$, $D_t=0$, and $\epsilon=0.2$.
In Movies 4 and 5, the dashed lines indicate that the four Gaussian lobes form a single probe; they do not contribute to its dynamics.

\begin{itemize}
\item Supplemental Movie 1: Spontaneous motion of a circular Gaussian probe.
The background shows $\Delta\rho(\br,V)/\rho_0=[\rho_{\mathrm{sim}}(\br,V)-\rho(\br,0)]/\rho_0$, where $\rho(\br,0)$ is the stationary reference density for the same probe at $V=0$.
The parameters are $N=5\times10^7$, $\Gamma=2.5$, and $\tilde M=10$.

\item Supplemental Movie 2: Spontaneous motion of an elliptical Gaussian probe.
The parameters are $N=10^7$, $\Gamma_g=1.6$, $\lambda=2$, $a_g=1$, and $\tilde M=10$.

\item Supplemental Movie 3: Direction selection of a fourfold Gaussian probe, shown in the probe frame.
The initial velocity is $v_x(0)=v_y(0)=0.5$, approximately the steady velocity at $\beta=2$.
During $t\in[10000,20000]$, $\beta$ increases linearly from 2 to 4; the velocity then relaxes toward the new fixed point [see Fig.~\ref{fig_4_fourfold_probe_multistability}].
The parameters are $N=10^7$, $\Gamma=6$, $a=1$, and $\tilde M=5$.

\item Supplemental Movie 4: Emergence of spontaneous motion of a fourfold Gaussian probe at $\beta=2$.
The parameters are $N=10^6$, $\Gamma=6$, $a=1$, and $\tilde M=10$.

\item Supplemental Movie 5: Emergence of spontaneous motion of a fourfold Gaussian probe at $\beta=4$.
The parameters are $N=10^6$, $\Gamma=6$, $a=1$, and $\tilde M=10$.
\end{itemize}

%% file: PRE/D_ai_statement.tex
\section{Use of Generative AI}

ChatGPT (GPT-5.6, OpenAI) was used during the research and verification process to assist in identifying relevant literature and exploring and checking analytical calculations.
All arguments, calculations, and cited sources included in the manuscript were independently verified by the authors. 
ChatGPT was also used for light editing, including grammar, wording, and language polishing.

%% file: main.bib
@article{pei2026transfer,
  author = {Pei, Ji-Hui and Maes, Christian},
  title = {Transfer of Active Motion from Medium to Probe via the Induced Friction and Noise},
  journal = {Phys. Rev. Lett.},
  volume = {136},
  pages = {038301},
  year = {2026},
  doi = {10.1103/bwk3-3nmn},
  url = {https://doi.org/10.1103/bwk3-3nmn}
}

@article{kim2024symmetry-breaking,
  author = {Kim, Ki-Won and Choe, Yunsik and Baek, Yongjoo},
  title = {Symmetry-breaking motility of penetrable objects in active fluids},
  journal = {Phys. Rev. E},
  volume = {109},
  number = {1},
  pages = {014614},
  year = {2024},
  month = jan,
  doi = {10.1103/PhysRevE.109.014614},
  url = {https://doi.org/10.1103/PhysRevE.109.014614}
}

@article{caprini2024emergent,
  author = {Caprini, Lorenzo and Ldov, Anton and Gupta, Rahul Kumar and Ellenberg, Hendrik and Wittmann, Ren{\'e} and L\"{o}wen, Hartmut and Scholz, Christian},
  title = {Emergent memory from tapping collisions in active granular matter},
  journal = {Commun. Phys.},
  volume = {7},
  number = {1},
  pages = {52},
  year = {2024},
  month = feb,
  doi = {10.1038/s42005-024-01540-w},
  url = {https://doi.org/10.1038/s42005-024-01540-w}
}

@article{granek2022anomalous,
  author = {Granek, Omer and Kafri, Yariv and Tailleur, Julien},
  title = {Anomalous Transport of Tracers in Active Baths},
  journal = {Phys. Rev. Lett.},
  volume = {129},
  number = {3},
  pages = {038001},
  year = {2022},
  month = jul,
  doi = {10.1103/PhysRevLett.129.038001},
  url = {https://doi.org/10.1103/PhysRevLett.129.038001}
}

@article{knippenberg2025negative,
  author = {Knippenberg, Timo and Bebon, Robin and Speck, Thomas and Bechinger, Clemens},
  title = {Negative Drag Force on Beating Flagellar-Shaped Bodies in Active Fluids},
  journal = {Phys. Rev. Lett.},
  volume = {135},
  number = {11},
  pages = {118301},
  year = {2025},
  month = sep,
  doi = {10.1103/3xbk-sxhh},
  url = {https://doi.org/10.1103/3xbk-sxhh}
}

@article{maes2020fluctuating,
  author = {Maes, Christian},
  title = {Fluctuating Motion in an Active Environment},
  journal = {Phys. Rev. Lett.},
  volume = {125},
  number = {20},
  pages = {208001},
  year = {2020},
  month = nov,
  doi = {10.1103/PhysRevLett.125.208001},
  url = {https://doi.org/10.1103/PhysRevLett.125.208001}
}

@article{dileonardo2010bacterial,
  author = {Di Leonardo, R. and Angelani, L. and Dell'Arciprete, D. and Ruocco, G. and Iebba, V. and Schippa, S. and Conte, M. P. and Mecarini, F. and De Angelis, F. and Di Fabrizio, E.},
  title = {Bacterial ratchet motors},
  journal = {Proc. Natl. Acad. Sci. U.S.A.},
  volume = {107},
  number = {21},
  pages = {9541--9545},
  year = {2010},
  month = may,
  doi = {10.1073/pnas.0910426107},
  url = {https://doi.org/10.1073/pnas.0910426107}
}

@article{sokolov2009swimming,
  author = {Sokolov, Andrey and Apodaca, Mario M. and Grzybowski, Bartosz A. and Aranson, Igor S.},
  title = {Swimming bacteria power microscopic gears},
  journal = {Proc. Natl. Acad. Sci. U.S.A.},
  volume = {107},
  number = {3},
  pages = {969--974},
  year = {2010},
  month = dec,
  doi = {10.1073/pnas.0913015107},
  url = {https://doi.org/10.1073/pnas.0913015107}
}

@article{mallory2014curvature-induced,
  author = {Mallory, S. A. and Valeriani, C. and Cacciuto, A.},
  title = {Curvature-induced activation of a passive tracer in an active bath},
  journal = {Phys. Rev. E},
  volume = {90},
  number = {3},
  pages = {032309},
  year = {2014},
  month = sep,
  doi = {10.1103/physreve.90.032309},
  url = {https://doi.org/10.1103/physreve.90.032309}
}

@article{baek2018generic,
  author = {Baek, Yongjoo and Solon, Alexandre P. and Xu, Xinpeng and Nikola, Nikolai and Kafri, Yariv},
  title = {Generic Long-Range Interactions Between Passive Bodies in an Active Fluid},
  journal = {Phys. Rev. Lett.},
  volume = {120},
  number = {5},
  pages = {058002},
  year = {2018},
  month = jan,
  doi = {10.1103/physrevlett.120.058002},
  url = {https://doi.org/10.1103/physrevlett.120.058002}
}

@article{wu2000particle,
  author = {Wu, Xiao-Lun and Libchaber, Albert},
  title = {Particle Diffusion in a Quasi-Two-Dimensional Bacterial Bath},
  journal = {Phys. Rev. Lett.},
  volume = {84},
  number = {13},
  pages = {3017--3020},
  year = {2000},
  month = mar,
  doi = {10.1103/PhysRevLett.84.3017},
  url = {https://doi.org/10.1103/PhysRevLett.84.3017}
}

@article{leptos2009dynamics,
  author = {Leptos, Kyriacos C. and Guasto, Jeffrey S. and Gollub, J. P. and Pesci, Adriana I. and Goldstein, Raymond E.},
  title = {Dynamics of Enhanced Tracer Diffusion in Suspensions of Swimming Eukaryotic Microorganisms},
  journal = {Phys. Rev. Lett.},
  volume = {103},
  number = {19},
  pages = {198103},
  year = {2009},
  month = nov,
  doi = {10.1103/PhysRevLett.103.198103},
  url = {https://doi.org/10.1103/PhysRevLett.103.198103}
}

@article{mino2011enhanced,
  author = {Mi{\~n}o, Gast{\'o}n and Mallouk, Thomas E. and Darnige, Thierry and Hoyos, Mauricio and Dauchet, Jeremi and Dunstan, Jocelyn and Soto, Rodrigo and Wang, Yang and Rousselet, Annie and Clement, Eric},
  title = {Enhanced Diffusion due to Active Swimmers at a Solid Surface},
  journal = {Phys. Rev. Lett.},
  volume = {106},
  number = {4},
  pages = {048102},
  year = {2011},
  month = jan,
  doi = {10.1103/PhysRevLett.106.048102},
  url = {https://doi.org/10.1103/PhysRevLett.106.048102}
}

@article{jepson2013enhanced,
  author = {Jepson, Alys and Martinez, Vincent A. and Schwarz-Linek, Jana and Morozov, Alexander and Poon, Wilson C. K.},
  title = {Enhanced diffusion of nonswimmers in a three-dimensional bath of motile bacteria},
  journal = {Phys. Rev. E},
  volume = {88},
  number = {4},
  pages = {041002(R)},
  year = {2013},
  month = oct,
  doi = {10.1103/PhysRevE.88.041002},
  url = {https://doi.org/10.1103/PhysRevE.88.041002}
}

@article{burkholder2017tracer,
  author = {Burkholder, Eric W. and Brady, John F.},
  title = {Tracer diffusion in active suspensions},
  journal = {Phys. Rev. E},
  volume = {95},
  number = {5},
  pages = {052605},
  year = {2017},
  month = may,
  doi = {10.1103/PhysRevE.95.052605},
  url = {https://doi.org/10.1103/PhysRevE.95.052605}
}

@article{xie2022activity-induced,
  author = {Xie, Chenliang and Liu, Yanan and Luo, Hao and Jing, Guangyin},
  title = {Activity-Induced Enhancement of Superdiffusive Transport in Bacterial Turbulence},
  journal = {Micromachines},
  volume = {13},
  number = {5},
  pages = {746},
  year = {2022},
  month = may,
  doi = {10.3390/mi13050746},
  url = {https://doi.org/10.3390/mi13050746}
}

@article{reichert2021tracer,
  author = {Reichert, Julian and Voigtmann, Thomas},
  title = {Tracer dynamics in crowded active-particle suspensions},
  journal = {Soft Matter},
  volume = {17},
  number = {46},
  pages = {10492--10504},
  year = {2021},
  doi = {10.1039/d1sm01092a},
  url = {https://doi.org/10.1039/d1sm01092a}
}

@article{grossmann2024non-gaussian,
  author = {Gro\ss{}mann, Robert and Bort, Lara S. and Moldenhawer, Ted and Stange, Maike and Panah, Setareh Sharifi and Metzler, Ralf and Beta, Carsten},
  title = {Non-{Gaussian} Displacements in Active Transport on a Carpet of Motile Cells},
  journal = {Phys. Rev. Lett.},
  volume = {132},
  number = {8},
  pages = {088301},
  year = {2024},
  month = feb,
  doi = {10.1103/PhysRevLett.132.088301},
  url = {https://doi.org/10.1103/PhysRevLett.132.088301}
}

@article{dhar2024active,
  author = {Dhar, Tanumoy and Saintillan, David},
  title = {Active transport of a passive colloid in a bath of run-and-tumble particles},
  journal = {Sci. Rep.},
  volume = {14},
  number = {1},
  pages = {11844},
  year = {2024},
  month = may,
  doi = {10.1038/s41598-024-62396-2},
  url = {https://doi.org/10.1038/s41598-024-62396-2}
}

@article{foffano2012colloids,
  author = {Foffano, G. and Lintuvuori, J. S. and Stratford, K. and Cates, M. E. and Marenduzzo, D.},
  title = {Colloids in Active Fluids: Anomalous Microrheology and Negative Drag},
  journal = {Phys. Rev. Lett.},
  volume = {109},
  number = {2},
  pages = {028103},
  year = {2012},
  month = jul,
  doi = {10.1103/PhysRevLett.109.028103},
  url = {https://doi.org/10.1103/PhysRevLett.109.028103}
}

@article{granek2020bodies,
  author = {Granek, Omer and Baek, Yongjoo and Kafri, Yariv and Solon, Alexandre P.},
  title = {Bodies in an interacting active fluid: far-field influence of a single body and interaction between two bodies},
  journal = {J. Stat. Mech.: Theory Exp.},
  volume = {2020},
  number = {6},
  pages = {063211},
  year = {2020},
  doi = {10.1088/1742-5468/ab7f34},
  url = {https://doi.org/10.1088/1742-5468/ab7f34}
}

@article{ni2015tunable,
  author = {Ni, Ran and Cohen Stuart, Martien A. and Bolhuis, Peter G.},
  title = {Tunable Long Range Forces Mediated by Self-Propelled Colloidal Hard Spheres},
  journal = {Phys. Rev. Lett.},
  volume = {114},
  number = {1},
  pages = {018302},
  year = {2015},
  month = jan,
  doi = {10.1103/PhysRevLett.114.018302},
  url = {https://doi.org/10.1103/PhysRevLett.114.018302}
}

@article{burkholder2020nonlinear,
  author = {Burkholder, Eric W. and Brady, John F.},
  title = {Nonlinear microrheology of active {Brownian} suspensions},
  journal = {Soft Matter},
  volume = {16},
  number = {4},
  pages = {1034--1046},
  year = {2020},
  doi = {10.1039/c9sm01713e},
  url = {https://doi.org/10.1039/c9sm01713e}
}

@article{peng2022forced,
  author = {Peng, Zhiwei and Brady, John F.},
  title = {Forced microrheology of active colloids},
  journal = {J. Rheol.},
  volume = {66},
  number = {5},
  pages = {955--972},
  year = {2022},
  month = aug,
  doi = {10.1122/8.0000504},
  url = {https://doi.org/10.1122/8.0000504}
}

@article{reichhardt2015active,
  author = {Reichhardt, C. and Olson Reichhardt, C. J.},
  title = {Active Microrheology in Active Matter Systems: Mobility, Intermittency, and Avalanches},
  journal = {Phys. Rev. E},
  volume = {91},
  number = {3},
  pages = {032313},
  year = {2015},
  month = mar,
  doi = {10.1103/PhysRevE.91.032313},
  url = {https://doi.org/10.1103/PhysRevE.91.032313}
}

@article{pei2025induced,
  author = {Pei, Ji-Hui and Maes, Christian},
  title = {Induced Friction on a Probe Moving in a Nonequilibrium Medium},
  journal = {Phys. Rev. E},
  volume = {111},
  number = {3},
  pages = {L032101},
  year = {2025},
  month = mar,
  doi = {10.1103/PhysRevE.111.L032101},
  url = {https://doi.org/10.1103/PhysRevE.111.L032101}
}

@article{knezevic2021oscillatory,
  author = {Kne{\v z}evi{\'c}, Milo{\v s} and Avil{\'e}s Podgurski, Luisa E. and Stark, Holger},
  title = {Oscillatory active microrheology of active suspensions},
  journal = {Sci. Rep.},
  volume = {11},
  number = {1},
  pages = {22706},
  year = {2021},
  month = nov,
  doi = {10.1038/s41598-021-02103-7},
  url = {https://doi.org/10.1038/s41598-021-02103-7}
}

@article{mori2020universal,
  author = {Mori, Francesco and Le Doussal, Pierre and Majumdar, Satya N. and Schehr, Gr\'egory},
  title = {Universal Survival Probability for a $d$-Dimensional Run-and-Tumble Particle},
  journal = {Phys. Rev. Lett.},
  volume = {124},
  number = {9},
  pages = {090603},
  year = {2020},
  month = mar,
  doi = {10.1103/PhysRevLett.124.090603},
  url = {https://doi.org/10.1103/PhysRevLett.124.090603}
}

@article{tailleur2008statistical,
  author = {Tailleur, J. and Cates, M. E.},
  title = {Statistical Mechanics of Interacting Run-and-Tumble Bacteria},
  journal = {Phys. Rev. Lett.},
  volume = {100},
  number = {21},
  pages = {218103},
  year = {2008},
  month = may,
  doi = {10.1103/PhysRevLett.100.218103},
  url = {https://doi.org/10.1103/PhysRevLett.100.218103}
}

@article{angelani2015run-and-tumble,
  author = {Angelani, Luca},
  title = {Run-and-tumble particles, telegrapher's equation and absorption problems with partially reflecting boundaries},
  journal = {J. Phys. A: Math. Theor.},
  volume = {48},
  number = {49},
  pages = {495003},
  year = {2015},
  month = nov,
  doi = {10.1088/1751-8113/48/49/495003},
  url = {https://doi.org/10.1088/1751-8113/48/49/495003}
}

@article{romanczuk2012active,
  author = {Romanczuk, P. and B\"{a}r, M. and Ebeling, W. and Lindner, B. and Schimansky-Geier, L.},
  title = {Active {Brownian} particles: From individual to collective stochastic dynamics},
  journal = {Eur. Phys. J. Spec. Top.},
  volume = {202},
  number = {1},
  pages = {1--162},
  year = {2012},
  month = mar,
  doi = {10.1140/epjst/e2012-01529-y},
  url = {https://doi.org/10.1140/epjst/e2012-01529-y}
}

@article{sevilla2015smoluchowski,
  author = {Sevilla, Francisco J. and Sandoval, Mario},
  title = {{Smoluchowski} diffusion equation for active {Brownian} swimmers},
  journal = {Phys. Rev. E},
  volume = {91},
  number = {5},
  pages = {052150},
  year = {2015},
  month = may,
  doi = {10.1103/PhysRevE.91.052150},
  url = {https://doi.org/10.1103/PhysRevE.91.052150}
}

@article{granek2024colloquium,
  author = {Granek, Omer and Kafri, Yariv and Kardar, Mehran and Ro, Sunghan and Tailleur, Julien and Solon, Alexandre},
  title = {Colloquium: Inclusions, Boundaries, and Disorder in Scalar Active Matter},
  journal = {Rev. Mod. Phys.},
  volume = {96},
  number = {3},
  pages = {031003},
  year = {2024},
  month = sep,
  doi = {10.1103/RevModPhys.96.031003},
  url = {https://doi.org/10.1103/RevModPhys.96.031003}
}

@article{maes2015friction,
  author = {Maes, Christian and Steffenoni, Stefano},
  title = {Friction and Noise for a Probe in a Nonequilibrium Fluid},
  journal = {Phys. Rev. E},
  volume = {91},
  number = {2},
  pages = {022128},
  year = {2015},
  month = feb,
  doi = {10.1103/PhysRevE.91.022128},
  url = {https://doi.org/10.1103/PhysRevE.91.022128}
}

@article{steffenoni2016interacting,
  author = {Steffenoni, Stefano and Kroy, Klaus and Falasco, Gianmaria},
  title = {Interacting {Brownian} Dynamics in a Nonequilibrium Particle Bath},
  journal = {Phys. Rev. E},
  volume = {94},
  number = {6},
  pages = {062139},
  year = {2016},
  month = dec,
  doi = {10.1103/PhysRevE.94.062139},
  url = {https://doi.org/10.1103/PhysRevE.94.062139}
}

@article{demery2014generalized,
  author = {D{\'e}mery, Vincent and B{\'e}nichou, Olivier and Jacquin, Hugo},
  title = {Generalized {Langevin} Equations for a Driven Tracer in Dense Soft Colloids: Construction and Applications},
  journal = {New J. Phys.},
  volume = {16},
  number = {5},
  pages = {053032},
  year = {2014},
  month = may,
  doi = {10.1088/1367-2630/16/5/053032},
  url = {https://doi.org/10.1088/1367-2630/16/5/053032}
}

@article{malakar2018steady,
  author = {Malakar, Kanaya and Jemseena, V. and Kundu, Anupam and Kumar, K. Vijay and Sabhapandit, Sanjib and Majumdar, Satya N. and Redner, S. and Dhar, Abhishek},
  title = {Steady State, Relaxation and First-Passage Properties of a Run-and-Tumble Particle in One Dimension},
  journal = {J. Stat. Mech.: Theory Exp.},
  volume = {2018},
  number = {4},
  pages = {043215},
  year = {2018},
  month = apr,
  doi = {10.1088/1742-5468/aab84f},
  url = {https://doi.org/10.1088/1742-5468/aab84f}
}

@article{cates2013when,
  author = {Cates, Michael E. and Tailleur, Julien},
  title = {When Are Active {Brownian} Particles and Run-and-Tumble Particles Equivalent? {C}onsequences for Motility-Induced Phase Separation},
  journal = {EPL},
  volume = {101},
  number = {2},
  pages = {20010},
  year = {2013},
  doi = {10.1209/0295-5075/101/20010},
  url = {https://doi.org/10.1209/0295-5075/101/20010}
}

@article{ramaswamy2010mechanics,
  author = {Ramaswamy, Sriram},
  title = {The Mechanics and Statistics of Active Matter},
  journal = {Annu. Rev. Condens. Matter Phys.},
  volume = {1},
  pages = {323--345},
  year = {2010},
  doi = {10.1146/annurev-conmatphys-070909-104101},
  url = {https://doi.org/10.1146/annurev-conmatphys-070909-104101}
}

@article{marchetti2013hydrodynamics,
  author = {Marchetti, M. C. and Joanny, J. F. and Ramaswamy, S. and Liverpool, T. B. and Prost, J. and Rao, Madan and Simha, R. Aditi},
  title = {Hydrodynamics of Soft Active Matter},
  journal = {Rev. Mod. Phys.},
  volume = {85},
  number = {3},
  pages = {1143--1189},
  year = {2013},
  month = jul,
  doi = {10.1103/RevModPhys.85.1143},
  url = {https://doi.org/10.1103/RevModPhys.85.1143}
}

@article{bechinger2016active,
  author = {Bechinger, Clemens and Di Leonardo, Roberto and L{\"o}wen, Hartmut and Reichhardt, Charles and Volpe, Giorgio and Volpe, Giovanni},
  title = {Active Particles in Complex and Crowded Environments},
  journal = {Rev. Mod. Phys.},
  volume = {88},
  number = {4},
  pages = {045006},
  year = {2016},
  month = nov,
  doi = {10.1103/RevModPhys.88.045006},
  url = {https://doi.org/10.1103/RevModPhys.88.045006}
}

@article{solon2015pressure,
  author = {Solon, Alexandre P. and Fily, Yaouen and Baskaran, Aparna and Cates, Michael E. and Kafri, Yariv and Kardar, Mehran and Tailleur, Julien},
  title = {Pressure Is Not a State Function for Generic Active Fluids},
  journal = {Nat. Phys.},
  volume = {11},
  number = {8},
  pages = {673--678},
  year = {2015},
  doi = {10.1038/nphys3377},
  url = {https://doi.org/10.1038/nphys3377}
}

@article{nikola2016active,
  author = {Nikola, Nikolai and Solon, Alexandre P. and Kafri, Yariv and Kardar, Mehran and Tailleur, Julien and Voituriez, Rapha{\"e}l},
  title = {Active Particles with Soft and Curved Walls: Equation of State, Ratchets, and Instabilities},
  journal = {Phys. Rev. Lett.},
  volume = {117},
  number = {9},
  pages = {098001},
  year = {2016},
  month = aug,
  doi = {10.1103/PhysRevLett.117.098001},
  url = {https://doi.org/10.1103/PhysRevLett.117.098001}
}

@article{peng2016diffusion,
  author = {Peng, Yi and Lai, Lipeng and Tai, Yi-Shu and Zhang, Kechun and Xu, Xinliang and Cheng, Xiang},
  title = {Diffusion of Ellipsoids in Bacterial Suspensions},
  journal = {Phys. Rev. Lett.},
  volume = {116},
  number = {6},
  pages = {068303},
  year = {2016},
  month = feb,
  doi = {10.1103/PhysRevLett.116.068303},
  url = {https://doi.org/10.1103/PhysRevLett.116.068303}
}

@article{yang2016dynamics,
  author = {Yang, Ou and Peng, Yi and Liu, Zhengyang and Tang, Chao and Xu, Xinliang and Cheng, Xiang},
  title = {Dynamics of Ellipsoidal Tracers in Swimming Algal Suspensions},
  journal = {Phys. Rev. E},
  volume = {94},
  number = {4},
  pages = {042601},
  year = {2016},
  month = oct,
  doi = {10.1103/PhysRevE.94.042601},
  url = {https://doi.org/10.1103/PhysRevE.94.042601}
}

@article{nordanger2022anisotropic,
  author = {Nordanger, Henrik and Morozov, Alexander and Stenhammar, Joakim},
  title = {Anisotropic Diffusion of Ellipsoidal Tracers in Microswimmer Suspensions},
  journal = {Phys. Rev. Fluids},
  volume = {7},
  number = {1},
  pages = {013103},
  year = {2022},
  month = jan,
  doi = {10.1103/PhysRevFluids.7.013103},
  url = {https://doi.org/10.1103/PhysRevFluids.7.013103}
}

@article{martens2012probability,
  author = {Martens, K. and Angelani, L. and Di Leonardo, R. and Bocquet, L.},
  title = {Probability distributions for the run-and-tumble bacterial dynamics: An analogy to the {Lorentz} model},
  journal = {Eur. Phys. J. E},
  volume = {35},
  number = {9},
  pages = {84},
  year = {2012},
  month = sep,
  doi = {10.1140/epje/i2012-12084-y},
  url = {https://doi.org/10.1140/epje/i2012-12084-y}
}

@article{dalcengio2019linear,
  author = {Dal Cengio, Sara and Levis, Demian and Pagonabarraga, Ignacio},
  title = {Linear Response Theory and {Green--Kubo} Relations for Active Matter},
  journal = {Phys. Rev. Lett.},
  volume = {123},
  pages = {238003},
  year = {2019},
  doi = {10.1103/PhysRevLett.123.238003},
  url = {https://doi.org/10.1103/PhysRevLett.123.238003}
}

@misc{kafri2026active,
  author = {Kafri, Yariv and Tailleur, Julien},
  title = {How active field theories couple to external potentials},
  year = {2026},
  eprint = {2603.22424},
  archivePrefix = {arXiv},
  primaryClass = {cond-mat.stat-mech}
}

@article{john2025progress,
  author = {John, Jacob and Panahi, Amirreza and Pu, Di and Natale, Giovanniantonio},
  title = {Progress in rheology of active colloidal systems},
  journal = {Curr. Opin. Colloid Interface Sci.},
  volume = {75},
  pages = {101886},
  year = {2025},
  doi = {10.1016/j.cocis.2024.101886},
  url = {https://doi.org/10.1016/j.cocis.2024.101886}
}

@article{burkholder2019fluctuation,
  author = {Burkholder, Eric W and Brady, John F},
  title = {Fluctuation-dissipation in active matter},
  journal = {J. Chem. Phys.},
  volume = {150},
  number = {18},
  pages = {184901},
  year = {2019},
  doi = {10.1063/1.5081725},
  url = {https://doi.org/10.1063/1.5081725}
}

@article{solon2022einstein,
  author = {Solon, Alexandre and Horowitz, Jordan M},
  title = {On the {Einstein} relation between mobility and diffusion coefficient in an active bath},
  journal = {J. Phys. A: Math. Theor.},
  volume = {55},
  number = {18},
  pages = {184002},
  year = {2022},
  doi = {10.1088/1751-8121/ac5d82},
  url = {https://doi.org/10.1088/1751-8121/ac5d82}
}

@article{shaebani2020computational,
  author = {Shaebani, M. Reza and Wysocki, Adam and Winkler, Roland G. and Gompper, Gerhard and Rieger, Heiko},
  title = {Computational models for active matter},
  journal = {Nat. Rev. Phys.},
  volume = {2},
  number = {4},
  pages = {181--199},
  year = {2020},
  doi = {10.1038/s42254-020-0152-1}
}

@article{lee2021geometry,
  author = {Lee, Jae Sung and Park, Jong-Min and Noh, Jae Dong and Park, Hyunggyu},
  title = {Geometry-induced rectification for an active object},
  journal = {Phys. Rev. Research},
  volume = {3},
  number = {4},
  pages = {L042011},
  year = {2021},
  doi = {10.1103/PhysRevResearch.3.L042011}
}

@article{vicsek1995novel,
  author = {Vicsek, Tamás and Czirók, András and Ben-Jacob, Eshel and Cohen, Inon and Shochet, Ofer},
  title = {Novel type of phase transition in a system of self-driven particles},
  journal = {Phys. Rev. Lett.},
  volume = {75},
  number = {6},
  pages = {1226--1229},
  year = {1995},
  doi = {10.1103/PhysRevLett.75.1226}
}

@article{kim2010multitime,
  author = {Kim, Kang and Saito, Shinji},
  title = {Multi-time density correlation functions in glass-forming liquids: Probing dynamical heterogeneity and its lifetime},
  journal = {J. Chem. Phys.},
  volume = {133},
  number = {4},
  pages = {044511},
  year = {2010},
  doi = {10.1063/1.3464331}
}

@article{squarcini2022spectral,
  author = {Squarcini, Alessio and Solon, Alexandre and Oshanin, Gleb},
  title = {Spectral density of individual trajectories of an active {B}rownian particle},
  journal = {New J. Phys.},
  volume = {24},
  number = {1},
  pages = {013018},
  year = {2022},
  doi = {10.1088/1367-2630/ac44e6}
}

@article{pietzonka2019autonomous,
  author = {Pietzonka, Patrick and Fodor, {\'E}tienne and Lohrmann, Christoph and Cates, Michael E. and Seifert, Udo},
  title = {Autonomous Engines Driven by Active Matter: Energetics and Design Principles},
  journal = {Phys. Rev. X},
  volume = {9},
  number = {4},
  pages = {041032},
  year = {2019},
  doi = {10.1103/PhysRevX.9.041032}
}

@misc{sm,
    note = {See Supplemental Material at [URL will be inserted by
publisher]}
}

@article{angelani2009selfstarting,
  author = {Angelani, Luca and Di Leonardo, Roberto and Ruocco, Giancarlo},
  title = {Self-Starting Micromotors in a Bacterial Bath},
  journal = {Phys. Rev. Lett.},
  volume = {102},
  number = {4},
  pages = {048104},
  year = {2009},
  doi = {10.1103/PhysRevLett.102.048104}
}

@article{pellicciotta2025wall,
  author = {Pellicciotta, Nicola and Bagal, Ojus Satish and Cannarsa, Maria Cristina and Bianchi, Silvio and Di Leonardo, Roberto},
  title = {Wall Torque Controls Propulsion of Curved Microstructures in Bacterial Baths},
  journal = {Phys. Rev. Lett.},
  volume = {135},
  number = {13},
  pages = {138302},
  year = {2025},
  doi = {10.1103/pnjv-6n5r}
}

@article{ning2023hydrodynamics,
  author = {Ning, Luhui and Lou, Xin and Ma, Qili and Yang, Yaochen and Luo, Nan and Chen, Ke and Meng, Fanlong and Zhou, Xin and Yang, Mingcheng and Peng, Yi},
  title = {Hydrodynamics-Induced Long-Range Attraction between Plates in Bacterial Suspensions},
  journal = {Phys. Rev. Lett.},
  volume = {131},
  number = {15},
  pages = {158301},
  year = {2023},
  doi = {10.1103/PhysRevLett.131.158301}
}

@article{jeanneret2016entrainment,
  author = {Jeanneret, Rapha{\"e}l and Pushkin, Dmitri O. and Kantsler, Vasily and Polin, Marco},
  title = {Entrainment dominates the interaction of microalgae with micron-sized objects},
  journal = {Nat. Commun.},
  volume = {7},
  number = {1},
  pages = {12518},
  year = {2016},
  doi = {10.1038/ncomms12518}
}

@article{davis2024active,
  author = {Davis, Luke K. and Proesmans, Karel and Fodor, {\'E}tienne},
  title = {Active Matter under Control: Insights from Response Theory},
  journal = {Phys. Rev. X},
  volume = {14},
  number = {1},
  pages = {011012},
  year = {2024},
  doi = {10.1103/PhysRevX.14.011012}
}

@article{bowick2022symmetry,
  author = {Bowick, Mark J. and Fakhri, Nikta and Marchetti, M. Cristina and Ramaswamy, Sriram},
  title = {Symmetry, Thermodynamics, and Topology in Active Matter},
  journal = {Phys. Rev. X},
  volume = {12},
  number = {1},
  pages = {010501},
  year = {2022},
  doi = {10.1103/PhysRevX.12.010501}
}

@article{olsonreichhardt2017ratchet,
  author = {Olson Reichhardt, C. J. and Reichhardt, C.},
  title = {Ratchet Effects in Active Matter Systems},
  journal = {Annu. Rev. Condens. Matter Phys.},
  volume = {8},
  number = {1},
  pages = {51--75},
  year = {2017},
  doi = {10.1146/annurev-conmatphys-031016-025522}
}

@article{takatori2014swim,
  author = {Takatori, Sho C. and Yan, Wen and Brady, John F.},
  title = {Swim Pressure: Stress Generation in Active Matter},
  journal = {Phys. Rev. Lett.},
  volume = {113},
  number = {2},
  pages = {028103},
  year = {2014},
  doi = {10.1103/PhysRevLett.113.028103}
}

@article{cates2015motility,
  author  = {M. E. Cates and J. Tailleur},
  title   = {Motility-induced phase separation},
  journal = {Annu. Rev. Condens. Matter Phys.},
  volume  = {6},
  pages   = {219},
  year    = {2015},
  doi     = {10.1146/annurev-conmatphys-031214-014710}
}

@article{toner1995long,
  author  = {J. Toner and Y. Tu},
  title   = {Long-range order in a two-dimensional dynamical {XY} model: How birds fly together},
  journal = {Phys. Rev. Lett.},
  volume  = {75},
  pages   = {4326},
  year    = {1995},
  doi     = {10.1103/PhysRevLett.75.4326}
}

@article{sanchez2012spontaneous,
  author  = {T. Sanchez and D. T. N. Chen and S. J. DeCamp and M. Heymann and Z. Dogic},
  title   = {Spontaneous motion in hierarchically assembled active matter},
  journal = {Nature},
  volume  = {491},
  pages   = {431},
  year    = {2012},
  doi     = {10.1038/nature11591}
}

@unpublished{companionPRL,
  author = {Woo, Chul-Ung and Choi, Jiwon and Rieger, Heiko},
  title = {Trajectory Statistics Govern Mechanical Power Transfer in Active Baths},
  note = {Companion manuscript},
  year = {2026}
}
